\documentstyle{mn}
\input{psfig.sty}
\newcommand{\etal}{{et al.~}}

\newcommand{\kmsmpc}{\>{\rm km}\,{\rm s}^{-1}\,{\rm Mpc}^{-1}}
\newcommand{\kms}{\>{\rm km}\,{\rm s}^{-1}}

\newcommand{\Mpc}{\>{\rm Mpc}}

\newcommand{\Msun}{\>{\rm M_{\odot}}}
\newcommand{\Lsun}{\>{\rm L_{\odot}}}
\newcommand{\MLsun}{\>({\rm M}/{\rm L})_{\odot}}

\newcommand{\Msunh}{\>h^{-1}\rm M_\odot}

\newcommand{\walpha}{\tilde{\alpha}}
\newcommand{\wLstar}{\tilde{L}^{*}}

\def\gtsima{$\; \buildrel > \over \sim \;$}
\def\ltsima{$\; \buildrel < \over \sim \;$}
\def\prosima{$\; \buildrel \propto \over \sim \;$}
\def\gsim{\lower.7ex\hbox{\gtsima}}
\def\lsim{\lower.7ex\hbox{\ltsima}}
\def\simgt{\lower.7ex\hbox{\gtsima}}
\def\simlt{\lower.7ex\hbox{\ltsima}}
\def\simpr{\lower.7ex\hbox{\prosima}}
\def\la{\lsim}
\def\ga{\gsim}
\def\lta{\la}
\def\gta{\ga}

\newcommand{\apj}{ApJ}
\newcommand{\apjs}{ApJS}
\newcommand{\aj}{AJ}
\newcommand{\mnras}{MNRAS}
\newcommand{\aap}{A\&A}

\newcommand{\pasp}{PASP}

\newdimen\hssize
\newdimen\hdsize
\begin{document}


\title[Linking Early and Late Type Galaxies to their Dark Matter Haloes]
      {Linking Early and Late Type Galaxies to their Dark Matter Haloes}
\author[van den Bosch, Yang \& Mo]
       {Frank C. van den Bosch$^{1}$, Xiaohu Yang$^{1,2}$, and H.J. Mo$^{1}$
        \thanks{E-mail: vdbosch@mpa-garching.mpg.de}\\
        $^1$Max-Planck-Institut f\"ur Astrophysik, Karl Schwarzschild
         Str. 1, Postfach 1317, 85741 Garching, Germany\\
        $^2$Center for Astrophysics, University of Science and Technology
         of China, Hefei, Anhui 230026, China}


\date{Accepted ........
      Received .......;
      in original form .......}


\maketitle

\label{firstpage}


\begin{abstract}
   Using data from the 2  degree Field Galaxy Redshift Survey (2dFGRS)
   we compute  the conditional  luminosity functions (CLFs)  of early-
   and late-type galaxies. These  functions give the average number of
   galaxies with  luminosities in  the range $L  \pm {\rm  d}L/2$ that
   reside in a  halo of mass $M$, and are  a powerful statistical tool
   to link the distribution of galaxies to that of dark matter haloes.
   Although  some amount  of  degeneracy remains,  the  CLFs are  well
   constrained.  They  indicate that the  average mass-to-light ratios
   of dark matter haloes have a  minimum of $\sim 100 h \MLsun$ around
   a halo mass of $\sim 3 \times 10^{11} h^{-1} \Msun$.  Towards lower
   masses $\langle  M/L \rangle$  increases rapidly, and  matching the
   faint-end slope  of the observed luminosity  function (LF) requires
   that haloes with $M < 10^{10} h^{-1} \Msun$ are virtually devoid of
   galaxies.  At the high mass end, the observed clustering properties
   of  galaxies require  that clusters  have  $b_J$-band mass-to-light
   ratios in  the range $500 -  1000 \, h \MLsun$.   Finally, the fact
   that early-type galaxies are more strongly clustered than late-type
   galaxies  requires that  the fraction  of late-type  galaxies  is a
   strongly  declining function  of halo  mass.  We  compute two-point
   correlation  functions as  function of  both luminosity  and galaxy
   type.  The  agreement with observations, in  terms of normalization
   and power-law slope, is remarkably good. When including predictions
   for  the correlation  functions of  faint galaxies  we find  a weak
   (strong)   luminosity  dependence   for  the   late   (early)  type
   galaxies. We also investigate the inferred halo occupation numbers.
   Late-type  and  faint  galaxies   reveal  a  shallower  $\langle  N
   \rangle(M)$  than bright, early-type  galaxies, which  explains why
   $\langle  N  \rangle(M)$ transforms  from  a  single power-law  for
   bright galaxies  to a more  complicated form when  fainter galaxies
   are included.   Finally we compare  our CLFs with  predictions from
   several semi-analytical  models for  galaxy formation.  As  long as
   these  models accurately  fit  the 2dFGRS  luminosity function  the
   agreement with our predictions  is remarkably good.  This indicates
   that  the   technique  used   here  has  recovered   a  statistical
   description of  how galaxies populate  dark matter haloes  which is
   not only in perfect agreement  with the data, but which in addition
   fits nicely within the standard framework for galaxy formation.
\end{abstract}




\section{Introduction}
\label{sec:intro}

According  to the  current paradigm  galaxies form  and  reside inside
extended cold dark matter (CDM) haloes. One of the ultimate challenges
in astrophysics is to obtain  a detailed understanding of how galaxies
with different properties occupy haloes of different masses. This link
between  galaxies and  dark matter  haloes  is an  imprint of  various
complicated physical  process related to galaxy formation  such as gas
cooling, star  formation, merging, tidal stripping and  heating, and a
variety  of feedback  processes. 

One  method  to  investigate  the  galaxy-dark  matter  connection  is
therefore  to consider {\it  ab initio}  models for  galaxy formation,
using either numerical simulations  (e.g., Katz, Weinberg \& Hernquist
1996; Fardal  \etal 2001;  Pearce \etal 2000;  Kay \etal  2002) and/or
``semi-analytical'' models (e.g., White \& Rees 1978; Kauffmann, White
\& Guiderdoni 1993; Somerville  \& Primack 1999; Kauffmann \etal 1999;
Cole \etal 2000; Benson \etal 2002; van den Bosch 2002). A downside of
this approach, however, is  that phenomenological descriptions have to
be used to describe a variety of poorly understood physical processes.
Consequently,   an  alternative  method   has  been   developed  which
completely  sidesteps  the   uncertainties  related  to  how  galaxies
form. This  method tries to infer  the link between  galaxies and dark
matter  haloes directly  from  the observed  clustering properties  of
galaxies.  Since haloes of  different mass  and galaxies  of different
luminosity and  type are  all clustered differently,  there is  only a
limited amount  of possibilities by which one  can distribute galaxies
over  dark matter  haloes such  that their  clustering  properties are
consistent with  observations.  The backbone  of this approach  is the
so-called halo model, which  views the evolved, non-linear dark matter
distribution  in  terms  of  its  halo building  blocks:  on  strongly
non-linear scales the dark matter  distribution is given by the actual
density  distributions  of the  virialized  haloes,  while on  larger,
close-to-linear scales,  the dark matter distribution is  given by the
distribution  of virialized  haloes (see  Cooray \&  Sheth 2002  for a
detailed review).  This halo model  has become more and  more accurate
due  to  the  fact  that  detailed  analytical  descriptions  for  the
structure and  clustering of dark matter haloes  have become available
(e.g., Navarro, Frenk  \& White 1997; Moore \etal  1998; Bullock \etal
2001; Mo \& White 1996, 2002; Power \etal 2002).

The halo model  can be naturally extended to address  the bias of {\it
galaxies}  by introducing  a model  for the  halo  occupation numbers,
$\langle N(M)  \rangle$, which describes how many  galaxies on average
(with  luminosities $L  > L_{\rm  min}$) occupy  a halo  of  mass $M$.
Numerous studies  in the past  have used these halo  occupation number
models to investigate how changes  in $\langle N(M) \rangle$ impact on
several statistical properties of the galaxy distribution, such as the
real-space  two-  and  three-point  correlation functions,  the  power
spectrum and bispectrum of galaxies, the galaxy-mass cross correlation
function,  the  pair-wise velocity  dispersions,  etc.  (Seljak  2000;
Scoccimarro \etal 2001; White 2001; Berlind \& Weinberg 2002; Scranton
2002a; Kang \etal 2002; Marinoni \& Hudson 2002; Kochanek \etal 2002).
In addition, several studies have confronted these models with data to
put  constraints on  $\langle N(M)  \rangle$ (Jing,  Mo  \& B\"{o}rner
1998; Peacock \&  Smith 2000; Marinoni \& Hudson  2002; Kochanek \etal
2002;  Jing, B\"orner \&  Suto 2002;  Bullock, Wechsler  \& Somerville
2002).  With large galaxy redshift surveys becoming available, such as
the Two-Degree Field Galaxy Redshift Survey (2dFGRS; see Colless \etal
2001) and the  Sloan Digital Sky  Survey (SDSS, see York  \etal 2000),
these  models   can  now  be  confronted  with   statistical  data  of
unprecedented  quality to  obtain  stringent constraints  on the  halo
occupation numbers, and therewith  on both cosmological parameters and
galaxy formation models.

In a recent paper, Yang, Mo \& van den Bosch (2002; hereafter Paper~1)
have taken this approach one  step further by considering the derivate
of  $\langle  N(M)  \rangle$   with  respect  to  $L_{\rm  min}$.   In
particular,  they  introduced   the  conditional  luminosity  function
(hereafter CLF)  $\Phi(L \vert M)  {\rm d}L$, which gives  the average
number of galaxies  with luminosities in the range  $L \pm {\rm d}L/2$
that reside in haloes of mass  $M$.  The advantage of the CLF over the
halo occupation function $\langle N(M)  \rangle$ is that it allows one
to address the  clustering properties of galaxies {\it  as function of
luminosity}.  In  addition, the CLF  yields a direct link  between the
halo mass  function $n(M)  {\rm d}M$, which  gives the number  of dark
matter haloes per comoving volume with masses in the range $M \pm {\rm
d}M/2$,  and the  galaxy luminosity  function (hereafter  LF) $\Phi(L)
{\rm d}L$, which gives the number of galaxies per comoving volume with
luminosities in the range $L\pm {\rm d}L/2$, according to
\begin{equation}
\label{phiL}
\Phi(L) = \int_{0}^{\infty} \Phi(L \vert M) \, n(M) \, {\rm d}M.
\end{equation}
Therefore, $\Phi(L \vert M)$ is not only constrained by the clustering
properties of  galaxies, as is the  case with $\langle  N(M) \rangle$,
but also  by   the observed galaxy   luminosity function. Furthermore,
knowledge  of $\Phi(L \vert  M)  {\rm d}L$  allows  one to compute the
average total luminosity of galaxies in a halo of mass $M$,
\begin{equation}
\label{LofM}
\langle L \rangle(M) = \int_{0}^{\infty} \Phi(L  \vert M) \, L \, {\rm
d}L
\end{equation}
and thus  the average  mass-to-light ratio as  function of  halo mass.
This $\langle  M/L \rangle(M)$ yields important  constraints on galaxy
formation  models,  as  it  is  a  direct measure  of  the  halo  mass
dependence of the galaxy formation efficiency. 

In Paper~1 we focussed on the cosmology dependence of the CLF. In this
paper, we use  data from the 2dFGRS to compute the  CLFs of both early
and   late-type  galaxies   and   compare  our   results with  several
semi-analytical models  for galaxy formation.   It is well  known that
galaxies  of different morphological  types have  different luminosity
functions   and  different   clustering   properties.   For   example,
early-type  galaxies have  higher  characteristic luminosities  (e.g.,
Efstathiou, Ellis \&  Peterson 1988; Loveday \etal 1992)  and are more
strongly clustered (e.g., Willmer, Da Costa \& Pellegrini 1998; Zehavi
\etal 2002) than late-type galaxies.  The CLFs presented here allow an
interpretation of  these morphological  dependencies in terms  of halo
occupation  numbers,  while our  comparison  with the  semi-analytical
models links their physical origin to the framework  of galaxy
formation models.
 
Throughout this  paper we define  $M$ to be  the halo mass  inside the
radius $R_{180}$  inside of which  the average density is  $180$ times
the cosmic  mean density, and $h$  is the Hubble constant  in units of
$100 \kmsmpc$.

\section{Observational Constraints}
\label{sec:obs}

In paper~1  we used data from the  2dFGRS to constrain the  CLF of the
entire galaxy population. This  data set included over 110500 galaxies
with  $17.0 <  b_J  <  19.2$ and  $z<  0.25$.  In  this  paper we  are
interested in the CLFs of the early- and late-type galaxies, for which
we  need separate  LFs  and separate  measurements  of the  luminosity
dependence of the clustering properties.

Madgwick \etal  (2002) used a  principal component analysis  of galaxy
spectra   taken  from   the  2dFGRS   to  obtain   a   {\it  spectral}
classification scheme.  They introduced the parameter $\eta$, a linear
combination  of the two  most significant  principal components,  as a
galaxy type classification measure. As shown by Madgwick \etal (2002),
$\eta$  follows a  bimodal distribution  and can  be interpreted  as a
measure   for    the   current    star   formation   rate    in   each
galaxy. Furthermore $\eta$ is well correlated with {\it morphological}
type  (Madgwick 2002).  In  what follows  we adopt  the classification
suggested by Madgwick  \etal and classify galaxies with  $\eta < -1.4$
as `early-types'  and galaxies with $\eta \geq  -1.4$ as `late-types'.
We caution the reader that despite the good correlation between $\eta$
and  morphological  type, there  is  not  a one-to-one  correspondence
between  our description of  early- and  late-type galaxies  and those
obtained using a more morphological criterion.
\begin{figure*}
\centerline{\psfig{figure=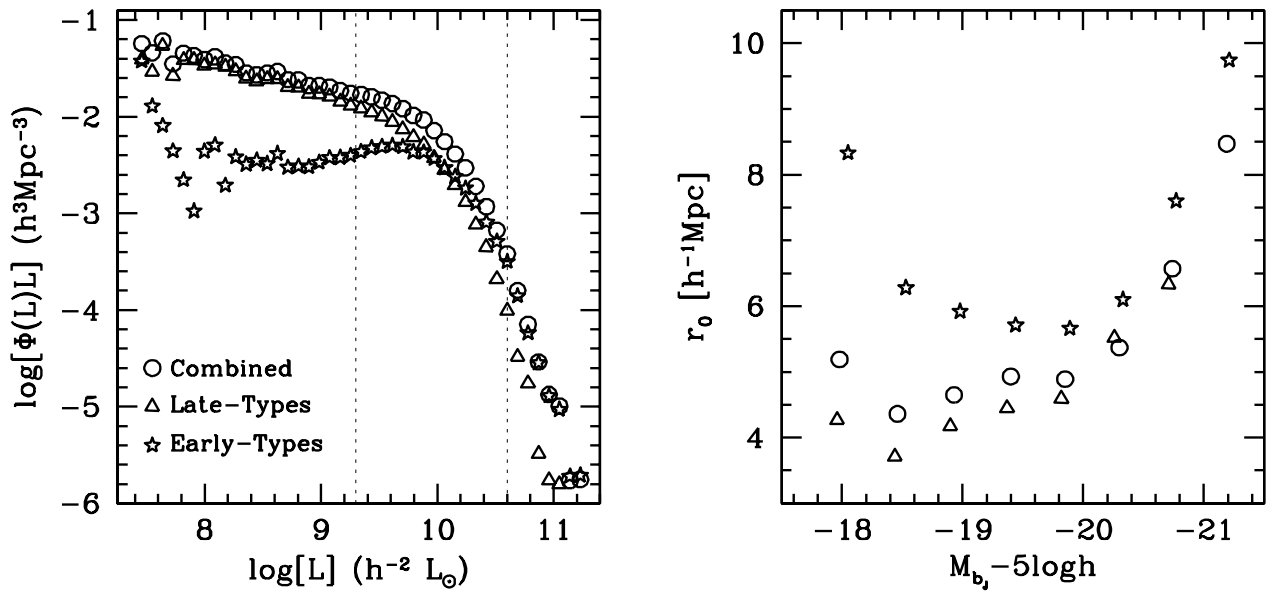,width=0.95\hdsize}}
\caption{The data used  to constrain the models. The  left panel plots
the LFs  of the early-type  galaxies (starred symbols),  the late-type
galaxies  (open triangles),  and  the combined  sample  of late-  plus
early-type  galaxies (open  circles).  For  clarity, no  errorbars are
shown  (but see Figure~\ref{fig:modelA}  below).  The  dotted vertical
lines  indicate  the  luminosity   range  over  which  measurement  of
clustering correlation lengths are available.  Right panel plots these
galaxy-galaxy  correlation  lengths as  function  of luminosity  (same
symbols);  Again,  for  clarity   no  errorbars  are  shown  (but  see
Figure~\ref{fig:modelA}).}
\label{fig:data}
\end{figure*}

\subsection{Luminosity Functions}
\label{sec:lf}

In order  to constrain  the CLFs of  early- and late-type  galaxies we
adopt the LFs in the photometric $b_J$ band computed by Madgwick \etal
(2002) from  the 2dFGRS.  This  sample is  restricted to  the redshift
range $z \leq  0.15$ and contains 75589 unique  galaxies with accurate
type-classification  based  on  $\eta$.  About  36  percent  of  these
galaxies have $\eta < -1.4$ and make up what we refer to as the sample
of  early-type   galaxies;  the  remainder  makes   up  the  late-type
galaxies. In what  follows, we shall refer to  the {\it entire} sample
(both late- and early-type galaxies) as the ``combined sample''.

The LFs obtained from this  sample of galaxies have been corrected for
completeness    effects,   and    a    self-consistent   method    for
$k$-corrections,  based   on  the  observed  2dF   spectra,  has  been
applied.  A cosmology  with $\Omega_0=0.3$  and $\Omega_{\Lambda}=0.7$
has  been   adopted,  which  corresponds  to   the  same  cosmological
parameters as we adopt throughout this paper.
  
The left panel of Figure~\ref{fig:data} plots the LFs for the combined
sample  (circles),  as  well   as  its  contributions  from  late-type
(triangles)  and  early-type  (starred  symbols) galaxies.   In  their
paper,  Madgwick \etal  (2002)  subdivided the  late-type galaxies  in
three sub-types.   The late-type LF  used here is computed  by summing
the  LFs  of  these  three  subtypes,  and by  adding  the  errors  in
quadrature. For clarity, no errorbars have been plotted, but these can
be seen in Figure~\ref{fig:modelA} below.  The main characteristics of
these LFs are  that (i) the LFs of the  combined and late-type samples
are well fit by a Schechter (1976) function (see Madgwick \etal (2002)
for the  parameters), (ii) early-  (late-) type galaxies  dominate the
total LF at the bright (faint) end, and (iii) the LF of the early-type
galaxies  reveals a  remarkable, though  only  marginally significant,
upturn at the faint-end.
  
\subsection{Correlation Lengths}
\label{sec:corr}

It is  straightforward to see  that there is  an infinite set  of CLFs
that,  given a  halo  mass  function, can  reproduce  the observed  LF
$\Phi_{\rm  obs}(L)$ of  galaxies.   For example,  all  CLFs given  by
$\Phi(L  \vert  M)  = [\Phi_{\rm  obs}(L)/n(M_0)]  \delta_{k}(M-M_0)$,
where $\delta_{k}(x)$ is the Kronecker  delta function and $M_0$ is an
arbitrary mass, yield exactly  the same LF, namely $\Phi_{\rm obs}(L)$
(see  equation~[\ref{phiL}]).  Therefore,  in order  to  constrain the
CLF,  additional constraints  are required.   Since more  massive dark
matter haloes  are more strongly  clustered (e.g., Mo \&  White 1996),
galaxy-galaxy correlation lengths as function of luminosity, $r_0(L)$,
put important  constraints on  how galaxies of  different luminosities
have to  be distributed over  haloes of different masses.   Indeed, as
shown in Paper~1, a combination of an observed LF with measurements of
$r_0(L)$ allow the CLF to be well constrained.

Because of the shear size of  the 2dFGRS it is possible to compute the
clustering  properties of  galaxies  of different  spectral types  and
luminosities in representative  volume-limited samples.  Norberg \etal
(2002) obtained   real  space   galaxy-galaxy   two-point  correlation
functions $\xi_{\rm gg}(r)$ from the same sample of 2dFGRS galaxies as
that  used by  Madgwick \etal  (2002) for  the LFs.   For a  number of
volume-limited subsamples (specified by a range in absolute magnitude)
Norberg \etal computed  $\xi_{\rm gg}(r)$ (assuming $\Omega_0=0.3$ and
$\Omega_{\Lambda}=0.7$) separately for  the late-type, the early-type,
and the combined samples.  Each of these correlation functions is well
fit      by     a      simple      power-law     $\xi_{\rm      gg}(r)
=(r/r_0)^{\gamma}$. Although there is  no clear trend of $\gamma$ with
either luminosity  or spectral type  (see Section~\ref{sec:corrfunc}),
the correlation  lengths $r_0$ depend strongly on  both luminosity and
spectral   type.    This   is    shown   in   the   right   panel   of
Figure~\ref{fig:data}, which plots $r_0(L)$ for the late-type galaxies
(triangles),  the  early-type  galaxies  (starred  symbols),  and  the
combined sample (circles).  For clarity, no errorbars are plotted (but
see   Figure~\ref{fig:modelA}  below).    Two  trends   are  apparent:
typically  $r_0$  increases   with  luminosity  (i.e.,  more  luminous
galaxies are more strongly clustered) and early-type galaxies are more
strongly  clustered than  late-type galaxies  of the  same luminosity.
This type-dependence is particularly strong for the fainter galaxies.

Note  that  the  correlation  function  itself may  also  evolve  with
redshift, an effect that has  not been corrected for. For the redshift
range covered by the 2dFGRS data used here (i.e., $z \leq 0.15$), this
effect is  fairly small, but we  nevertheless take it  into account in
our modeling (see Section~\ref{sec:theory} below).

\section{Theoretical background}
\label{sec:theory}

Our main goal in this paper is to use the observed LFs and correlation
lengths  to  constrain  the  CLFs  of  the  early-type  and  late-type
galaxies. Here we briefly describe  how to compute LFs and correlation
functions from a  given CLF. More details can be  found in Paper~1 and
references therein.

To  compute the galaxy  LF $\Phi(L)$  from the  CLF $\Phi(L  \vert M)$
(equation~[\ref{phiL}])  one   needs  the  (cosmology-dependent)  mass
function $n(M)$ of dark matter haloes, which (at $z=0$) is given by
\begin{equation}
\label{halomf}
n(M) \, {\rm d}M = {\bar{\rho} \over M^2} \nu f(\nu) \,
\left| {{\rm d} {\rm ln} \sigma \over {\rm d} {\rm ln} M}\right|
{\rm d}M.
\end{equation}
Here $\bar{\rho}$ is the mean matter density of the Universe at $z=0$,
$\nu = \delta_c/\sigma(M)$,  $\delta_c$  is the critical   overdensity
required for collapse at $z=0$, $f(\nu)$ is a  function of $\nu$ to be
specified below, and $\sigma(M)$ is the linear rms mass fluctuation on
mass scale $M$, which is given by the linear power spectrum of density
perturbations $P(k)$ as
\begin{equation}
\label{variance}
\sigma^2(M) = {1 \over 2 \pi^2} \int_{0}^{\infty} P(k) \;
\widehat{W}_{M}^2(k) \; k^2 \; {\rm d}k,
\end{equation}
where $\widehat{W}_{M}(k)$  is the Fourier transform  of the smoothing
filter on mass scale $M$.

Throughout we  adopt the  form of $f(\nu)$  suggested by Sheth,  Mo \&
Tormen (2001):
\begin{equation}
\label{fnuST}
\nu \, f(\nu) = 0.644 \,\left(1 + {1\over \nu'{^{0.6}}}\right)\
\left({\nu'{^2}\over 2\pi}\right)^{1/2}
\exp\left(-{\nu'{^2} \over 2}\right)\,
\end{equation}
with $\nu'=0.841\,\nu$.  The resulting mass function has been shown to
be in excellent agreement with  numerical simulations, as long as halo
masses  are defined  as the  masses inside  a sphere  with  an average
overdensity of about $180$ (Sheth  \& Tormen 1999; Jenkins \etal 2001;
White  2002).  Therefore,  in what  follows we  consistently  use that
definition of halo  mass, and we use the CDM  power spectrum $P(k)$ of
Efstathiou, Bond  \& White  (1992) with a  spatial top-hat  filter for
which
\begin{equation}
\label{THfour}
\widehat{W}_{M}(k) = {3  \over (k R)^3} \left[ \sin(k R)  - k R \cos(k
R)\right]
\end{equation}
where the mass $M$ and filter radius $R$ are related according to $M
= 4 \pi \bar{\rho} R^3 / 3$.

In order to  compute $r_0(L)$ from the CLF we  proceed as follows.  In
the halo model  it is natural to consider  the galaxy-galaxy two-point
correlation  function, $\xi_{\rm  gg}(r)$,  to be  build  up from  two
parts; a 1-halo term,  $\xi^{1 {\rm h}}_{\rm gg}(r)$, which represents
the correlation due  to pairs of galaxies within the  same halo, and a
2-halo   term,   $\xi^{2  {\rm   h}}_{\rm   gg}(r)$,  describing   the
contribution due  to pairs of  galaxies that occupy  different haloes.
The observed correlation lengths are all well in excess of $3.5 h^{-1}
\Mpc$.  At radii  this large the contribution from  the 1-halo term is
negligible and for the purpose of calculating correlation lengths only
the 2-halo term is required.

As we  show in  Appendix~A, the 2-halo  term of $\xi_{\rm  gg}(r)$ for
galaxies with $L_1 < L < L_2$ can be written as
\begin{equation}
\label{xitwoh}
\xi_{\rm gg}^{\rm 2h}(r) = \overline{b}^2 \, \xi_{\rm dm}^{\rm 2 h}(r) ,
\end{equation}
with $\xi_{\rm  dm}^{\rm 2h}(r)$  the 2-halo term  of the  dark matter
mass correlation function at $z=0$, which at scales of the correlation
length is well fit by a single power-law
\begin{equation}
\label{xipowlaw}
\xi_{\rm  dm}^{\rm 2h}(r) \simeq \xi_{\rm dm}(r) \simeq
\left( {r \over r_{0,{\rm dm}}} \right)^{-1.75}
\end{equation}
and
\begin{equation}
\label{averbias}
\overline{b} = {
\int_{0}^{\infty} n(M) \, \langle N(M) \rangle \, b(M) \, {\rm d}M 
\over
\int_{0}^{\infty} n(M) \, \langle N(M)\rangle \, {\rm d}M}
\end{equation}
Here
\begin{equation}
\label{nlm}
\langle N(M) \rangle = \int_{L_1}^{L_2} \Phi (L\vert M) \, {\rm d}L
\end{equation}
is the mean  number of galaxies in the  specified luminosity range for
haloes of  mass $M$, and $b(M)$ is  the bias of dark  matter haloes of
mass  $M$ with  respect  to  the dark  matter  mass distribution  (see
Appendix~A).   Note   that  the  2-halo  term   of  the  galaxy-galaxy
correlation function is  completely specified by the CLF  and does not
require knowledge about how galaxies are distributed inside individual
dark  matter haloes.  

Since   $\overline{b}$   is    scale-independent,   we   can   rewrite
equation~(\ref{xitwoh})  directly in terms  of the  correlation length
for galaxies as
\begin{equation}
\label{r0Lgal}
r_0 = \overline{b}^{1.143} \, r_{0,{\rm dm}}
\end{equation}

There  is one additional  effect, though,  that we  have to  take into
account. The  correlation functions  obtained by Norberg  \etal (2002)
have not been corrected for possible redshift evolution.  This implies
that the $r_0(L)$  measurements do not correspond to  $z=0$.  In fact,
since more luminous  galaxies can be detected out  to higher redshift,
the  correlation lengths  of  brighter galaxies  correspond to  galaxy
populations  with a higher  median redshift.   Therefore, in  order to
compare model and observations in a consistent way, we have to compute
the correlation  lengths at the characteristic redshift  of the sample
in   consideration.   We   take   this  into   account  by   replacing
$\overline{b}$  in   equation~(\ref{r0Lgal})  with  $\overline{b}_{\rm
eff}(\overline{z})$.  Here $\overline{z}$ is  the mean redshift of the
galaxies  in  consideration  (taken   from  Norberg  \etal  2002)  and
$\overline{b}_{\rm   eff}$   is   the   effective  bias   defined   in
Appendix~B. Since  the 2dFGRS sample used  here is limited  to $z \leq
0.15$, this redshift  correction is only modest, amounting  to no more
than a few percent change in $r_0$.
  
\section{Modeling the conditional luminosity function}
\label{sec:condLF}

In Paper~1 we adopted a particular parameterization of the CLF for the
entire galaxy population (early plus  late type galaxies), and we used
a $\chi^2$ minimization routine to find those parameters that best fit
the observed  $\Phi(L)$ and $r_0(L)$.   Here we seek to  constrain two
independent CLFs,  namely that  of the early-type  galaxies, hereafter
$\Phi_e(L  \vert M)$, and  that of  the late-type  galaxies, hereafter
$\Phi_l(L   \vert  M)$.    In  principle   we  could   use   the  same
parameterizations as in Paper~1  for both of these CLFs independently.
However, there  is an  additional constraint that  these CLFs  have to
fulfill:  their sum  must be  equal to  the CLF  of the  entire galaxy
population, and thus be consistent  with the $\Phi(L)$ and $r_0(L)$ of
the combined sample.  We  therefore use a slightly different, two-step
method which automatically obeys  the constraint that $\Phi(L \vert M)
=  \Phi_l(L \vert  M) +  \Phi_e(L \vert  M)$.  We  first use  the same
method  as in  Paper~1 to  obtain  the CLF  $\Phi(L \vert  M)$ of  the
combined  sample   (step  one).    Next  we  introduce   the  function
$f_l(L,M)$, which  specifies the  late-type fraction of  galaxies with
luminosity  $L$  in  haloes  of  mass  $M$.  The  CLFs  of  late-  and
early-type galaxies are then given by
\begin{equation}
\label{CLFl}
\Phi_{l}(L \vert M) {\rm d}L = f_l(L,M) \, \Phi(L \vert M) {\rm d}L
\end{equation}
and, by definition,
\begin{equation}
\label{CLFe}
\Phi_{e}(L \vert M) \, {\rm d}L = \left[ 1 - f_l(L,M) \right] \,
\Phi(L \vert M) \, {\rm d}L 
\end{equation}
What  remains (step two)  is to  find the  $f_l(L,M)$ that,  given our
best-fit  CLF for  the  combined  population, best  fits  the LFs  and
correlation lengths of the early- and late-type galaxies.

Following Paper~1 we assume that the CLF of the combined sample can be
described by a Schechter function:
\begin{equation}
\label{phiLM}
\Phi(L \vert M) {\rm d}L = {\tilde{\Phi}^{*} \over \wLstar} \,
\left({L \over \wLstar}\right)^{\walpha} \,
\, {\rm exp}(-L/\wLstar) \, {\rm d}L
\end{equation}
Here   $\wLstar   =    \wLstar(M)$,   $\walpha   =   \walpha(M)$   and
$\tilde{\Phi}^{*} =  \tilde{\Phi}^{*}(M)$; i.e., the  three parameters
that describe the conditional LF depend on $M$.  In what follows we do
not explicitly write this  mass dependence, but consider it understood
that quantities with  a tilde are functions of $M$.  

We adopt  the same parameterizations  of these three parameters  as in
Paper~1, which we repeat  here for completeness. Readers interested in
the  motivations  behind  these  particular choices  are  referred  to
Paper~1. For  the total mass-to-light ratio  of a halo of  mass $M$ we
write
\begin{equation}
\label{MtoLmodel}
\left\langle {M \over L} \right\rangle (M) = {1 \over 2} \,
\left({M \over L}\right)_0
\left[ \left({M \over M_c}\right)^{-\gamma_1} +
\left({M \over M_c}\right)^{\gamma_2}\right],
\end{equation}
which has four free parameters: a characteristic mass $M_c$, for which
the  mass-to-light ratio is  equal   to  $(M/L)_0$,  and two   slopes,
$\gamma_1$ and $\gamma_2$, which  specify the behavior of $\langle M/L
\rangle$ at  the  low and high   mass ends,  respectively.   A similar
parameterization is used for the characteristic luminosity $\wLstar$:
\begin{equation}
\label{LstarM}
{M \over \wLstar(M)} = {1 \over 2} \, \left({M \over L}\right)_0 \,
f(\walpha) \, \left[ \left({M \over M_c}\right)^{-\gamma_1} +
\left({M \over M_2}\right)^{\gamma_3}\right].
\end{equation}
Here
\begin{equation}
\label{falpha}
f(\walpha) = {\Gamma(\walpha+2) \over \Gamma(\walpha+1,1)}.
\end{equation}
with $\Gamma(x)$  the Gamma function and  $\Gamma(a,x)$ the incomplete
Gamma  function.   This   parameterization  has  two  additional  free
parameters:  a  characteristic  mass   $M_2$  and  a  power-law  slope
$\gamma_3$.  For $\walpha(M)$ we adopt:
\begin{equation}
\label{alphaM}
\walpha(M) = \alpha_{15} + \zeta \, \log(M_{15}).
\end{equation}
Here   $M_{15}$  is  the halo  mass   in   units  of $10^{15} \Msunh$,
$\alpha_{15} = \walpha(M_{15}=1)$, and $\zeta$ describes the change of
the    faint-end slope $\walpha$  with   halo  mass.   Note that  once
$\walpha$     and    $\wLstar$    are   given,    the    normalization
$\tilde{\Phi}^{*}$    of       the    CLF   is       obtained  through
equation~(\ref{MtoLmodel}), using the  fact  that the total  (average)
luminosity in a halo of mass $M$ is given by
\begin{equation}
\label{meanL}
\langle L \rangle(M) = \int_{0}^{\infty}  \Phi(L \vert M) \, L \, {\rm
d}L = \tilde{\Phi}^{*} \, \wLstar \, \Gamma(\walpha+2).
\end{equation}
Finally, we introduce the mass scale $M_{\rm min}$ below which the CLF
is zero;  i.e., we assume that no  stars form inside haloes  with $M <
M_{\rm min}$.   Motivated by reionization  considerations (see Paper~1
for details) we adopt $M_{\rm  min} = 10^{9} h^{-1} \Msun$ throughout.
This lower-mass  limit does  not significantly influence  our results.
For instance,  changing $M_{\rm min}$ to either  $10^{8} h^{-1} \Msun$
or $10^{10} h^{-1} \Msun$ has only a very modest impact on the results
presented below.

In order to split the CLF in early- and late-type galaxies we make the
assumption that $f_l(L,M)$ has a quasi-separable form
\begin{equation}
\label{fracdef}
f_l(L,M) = g(L) \, h(M) \, q(L,M)
\end{equation}
Here
\begin{equation}
\label{qlm}
q(L,M) = \left\{ \begin{array}{lll}
1                      & \mbox{if $g(L) \, h(M) \leq 1$} \\
{1 \over g(L) \, h(M)} & \mbox{if $g(L) \, h(M) > 1$}
\end{array} \right.
\end{equation}
is to ensure that $f_l(L,M) \leq 1$. For the $g(L)$ and $h(M)$ adopted
here (see below),  $q(L,M) = 1$ for the vast  majority of the relevant
$(L,M)$ parameter  space, and  therefore $f_l(L,M)$ can  be considered
separable  to good  accuracy.  This  basically implies  that  the {\it
shapes} (but not the absolute values) of the conditionals $f_l(L \vert
M)$ and $f_l(M \vert L)$ are independent of $M$ and $L$, respectively.
Note that  $g(L)$ is not the  same as the fraction  ${\cal F}_l(L)$ of
all galaxies with luminosity $L$  that are late-type, which instead is
given by
\begin{equation}
\label{calFl}
{\cal F}_l(L) = {\Phi_{l}(L) \over \Phi(L)} =  
{\int_{0}^{\infty} f_l(L,M) \, \Phi(L \vert M) \, n(M) \, {\rm d}M \over
 \int_{0}^{\infty} \Phi(L \vert M) \, n(M) \, {\rm d}M}
\end{equation}
Similarly, the fraction of all galaxies in haloes of mass $M$ that are
late-type is given by
\begin{equation}
\label{calFm}
{\cal F}_l(M) =  
{\int_{0}^{\infty} f_l(L,M) \, \Phi(L \vert M) \, {\rm d}L \over
 \int_{0}^{\infty} \Phi(L \vert M) \, {\rm d}L}
\end{equation}
The challenge now is to find the $g(L)$ and $h(M)$ that, given the CLF
for the  combined sample, reproduce  the observed LFs and  $r_0(L)$ of
the  early-  and late-type  galaxies.   To  achieve  this we  use  the
following estimate for $g(L)$:
\begin{equation}
\label{gl}
g(L) = {\hat{\Phi}_l(L) \over \hat{\Phi}(L)}  
{\int_{0}^{\infty} \Phi(L \vert M) \, n(M) \, {\rm d}M \over
 \int_{0}^{\infty} \Phi(L \vert M) \, h(M) \, n(M) \, {\rm d}M}
\end{equation}
where  $\hat{\Phi}_l(L)$ and  $\hat{\Phi}(L)$ correspond  to  the {\it
observed}  LFs   of  the   late-type  and  combined   galaxy  samples,
respectively. For any $L$ for which $f_l(L,M)$ is separable (i.e., for
which $q(L,M)  = 1$  for all $M$),  this implies  that the LFs  of the
early- and late-type galaxies are given by
\begin{equation}
\label{fitLFl}
\Phi_{e,l}(L) = \hat{\Phi}_{e,l}(L) \, {\Phi(L) \over \hat{\Phi}(L)},
\end{equation}
i.e., as long as the observed LF of the combined sample is well fit by
the  model, the  same  will be  true for  the  LFs of  the early-  and
late-type galaxies.  For the $L$ for  which $q(L,M) <  1$ a (typically
small) correction to~(\ref{fitLFl}) applies.   What remains is to find
the  $h(M)$  that best  reproduces  the  $r_0(L)$  of the  early-  and
late-type galaxies.  This requires the average biases $\overline{b}_l$
and $\overline{b}_e$ for  late- and early-type galaxies, respectively,
which are obtained  from equations~(\ref{averbias}) and~(\ref{nlm}) by
replacing  $\Phi(L \vert M)$  with $\Phi_l(L  \vert M)$  and $\Phi_e(L
\vert M)$, respectively.

After experimenting  with a variety of different  functional forms for
$h(M)$ we finally decided to adopt
\begin{equation}
\label{hm}
h(M) = \max \left( 0, \min\left[ 1, \left({{\rm log}(M/M_0) \over {\rm
log}(M_1/M_0)} \right) \right] \right)
\end{equation}
Here $M_0$ and $M_1$ are two free parameters, defined as the masses at
which $h(M)$ takes on the values $0$ and $1$, respectively.  Note that
both $h(M)$  and $g(L)$ can in  principle take on  any positive value.
However, for computational  reasons we limit $h(M)$ to  take values in
the range  $[0,1]$. This  does not impact  our results, as  any linear
scaling    of    $h(M)$   reflects    itself    in   $g(L)$    through
equation~(\ref{gl}).

Our  model  for the  three  CLFs  thus contains  a  total  of 10  free
parameters: 4  characteristic masses;  $M_c$, $M_0$, $M_1$  and $M_2$,
four   parameters   that   describe  the   various   mass-dependencies
$\gamma_1$, $\gamma_2$, $\gamma_3$  and $\zeta$, and 2 normalizations;
one for the mass-to-light ratio,  $(M/L)_0$, and one for the faint-end
slope of the CLF, $\alpha_{15}$.  Although this may seem an awful lot,
it  is  important  to  realize  that  we use  this  model  to  fit  67
independent data  points\footnote{Because of our  definition of $g(L)$
only  67  of   the  in  total  155  data   points  are  independent.}.
Furthermore,  as  we  show  in  Section~\ref{sec:res}  below  one  can
actually use  observational constraints to  fix several of  these free
parameters.   In  addition,  the  data  is of  sufficient  quality  to
constrain the  model freedom.  Alternatively,  we could have  chosen a
more restrictive (with  fewer free parameters) form for  the CLFs, but
lacking  both  observational  and  physical  motivations  for  a  more
preferred form of the CLF we felt the need to be sufficiently general.
On the  other side,  one might  argue that because  of this  lack, our
model is  actually too  constrained.  For instance,  the fact  that we
assume a Schechter function for the CLF is at best weakly motivated by
the observed  LFs of  clusters of galaxies  (e.g., Muriel,  Valotto \&
Lambas  1998;  Beijersbergen \etal  2002;  Trentham  \& Hodgkin  2002;
Mart{\'\i}nez \etal 2002).   However, as we show below,  our model can
accurately fit  all observations. Therefore,  the use of  more general
models would  only results in a  larger amount of  degeneracy and will
therefore  have to await  more stringent  constraints from  either the
SDSS or the completed 2dFGRS.
\begin{figure*}
\centerline{\psfig{figure=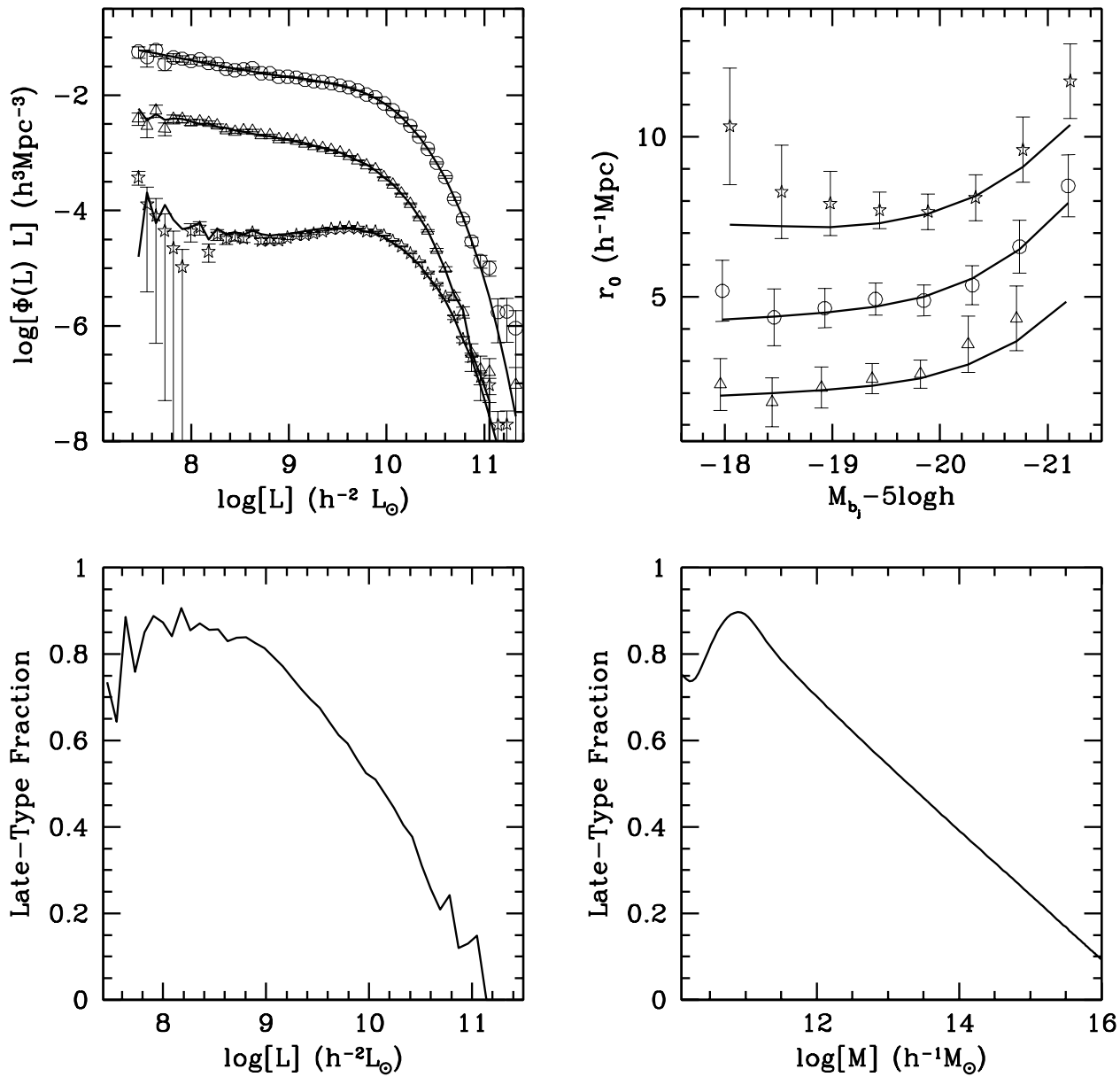,width=0.95\hdsize}}
\caption{Results for  model A. The upper  two panels plot the observed
LFs (left  panel) and correlation  lengths (right panel).  Symbols are
the same as in  Figure~\ref{fig:data}, while the solid lines  indicate
the model results. For clarity, the  LFs have been separated from each
other by   one  order of  magnitude  in  the   $y$-direction, with the
combined  LF untranslated. Similarly,  the  $r_0(L)$ of the early- and
late-type galaxies are  offset  by $+2 h^{-1}   \Mpc$ and  $-2  h^{-1}
\Mpc$, respectively.  Note  the good quality of  the fits.  The  lower
panels plot  the   fraction of  late-type  galaxies   as  function  of
luminosity (left) and halo mass (right).}
\label{fig:modelA}
\end{figure*}

\section{Results}
\label{sec:res}

The cosmological  model sets the  halo mass function $n(M)  {\rm d}M$,
the dark matter two-point  correlation function $\xi_{\rm dm}(r)$, and
the halo bias function $b(M)$.  Therefore, different cosmologies imply
different  halo occupancy  functions. In  this paper  we focus  on the
$\Lambda$CDM cosmology with  $\Omega_0=0.3$, $\Omega_{\Lambda} = 0.7$,
$h = 0.7$, $\Gamma =\Omega_0  \, h = 0.21$, and $\sigma_8=0.9$.  These
cosmological parameters are consistent with recent measurements of the
cosmic microwave  background (e.g., Pryke \etal 2001)  and large scale
structure  measurements  (e.g., Tegmark,  Hamilton  \&  Xu 2002).   In
addition, we  have shown  in Paper~1 that  this cosmology is  also the
most successful in explaining  the LF and galaxy clustering luminosity
dependence  observed.    Therefore,  in  what   follows,  we  restrict
ourselves to this ``concordance'' cosmology.

Having specified our  model, the  observational constraints,  and  the
cosmological   parameters,  we now proceed   as  follows. We first use
Powell's  multi-dimensional direction set   method (e.g., Press  \etal
1992) to find  the parameters of the CLF  of the combined  sample that
minimize
\begin{equation}
\label{chisq}
\chi^2 = {\chi^2(\Phi) \over N_{\Phi}} + {\chi^2(r_0) \over N_{r}}.
\end{equation}
Here
\begin{equation}
\label{chisqLF}
\chi^2(\Phi) = \sum_{i=1}^{N_{\Phi}}
\left[ {\Phi(L_i) - \hat{\Phi}(L_i) \over \Delta \hat{\Phi}(L_i)} \right]^2,
\end{equation}
and
\begin{equation}
\label{chisqr0}
\chi^2(r_0) = \sum_{i=1}^{N_{r}}
\left[ {r_0(L_i) - \hat{r}_0(L_i) \over \Delta \hat{r}_0(L_i)} \right]^2,
\end{equation}
with $\hat{\Phi}(L_i)$  and $\hat{r}_0(L_i)$ the  observed values, and
$N_{\Phi}=44$ and $N_{r}=8$ the number of corresponding data points.

Note that the scaling of  $\chi^2$ with $N_{\Phi}$ and $N_{r}$ implies
that  we assign  equal  weights to  $\chi^2(\Phi)$ and  $\chi^2(r_0)$.
This  differs from  the  proper statistical  $\chi^2$  for which  each
individual {\it  measurement} should receive  equal weights.  However,
since $N_{\Phi} >  N_{r}$ and since the errors  on $\hat{\Phi}(L)$ are
typically much smaller than the errors on $\hat{r}_0(L)$, the $\chi^2$
minimization routine  would give  much more weight  to fitting  the LF
than  to  fitting  the  correlation  lengths.  Note  that  the  proper
$\chi^2$ is only well defined if there are no systematic errors in the
measurements  and  if  the  error properties  of  $\hat{\Phi}(L)$  and
$\hat{r}_0(L)$ are  the same.  It is unlikely  that these requirements
are  fulfilled,  and  our   modified  $\chi^2$  is  therefore  equally
meaningful as that of the proper $\chi^2$.
\begin{figure*}
\centerline{\psfig{figure=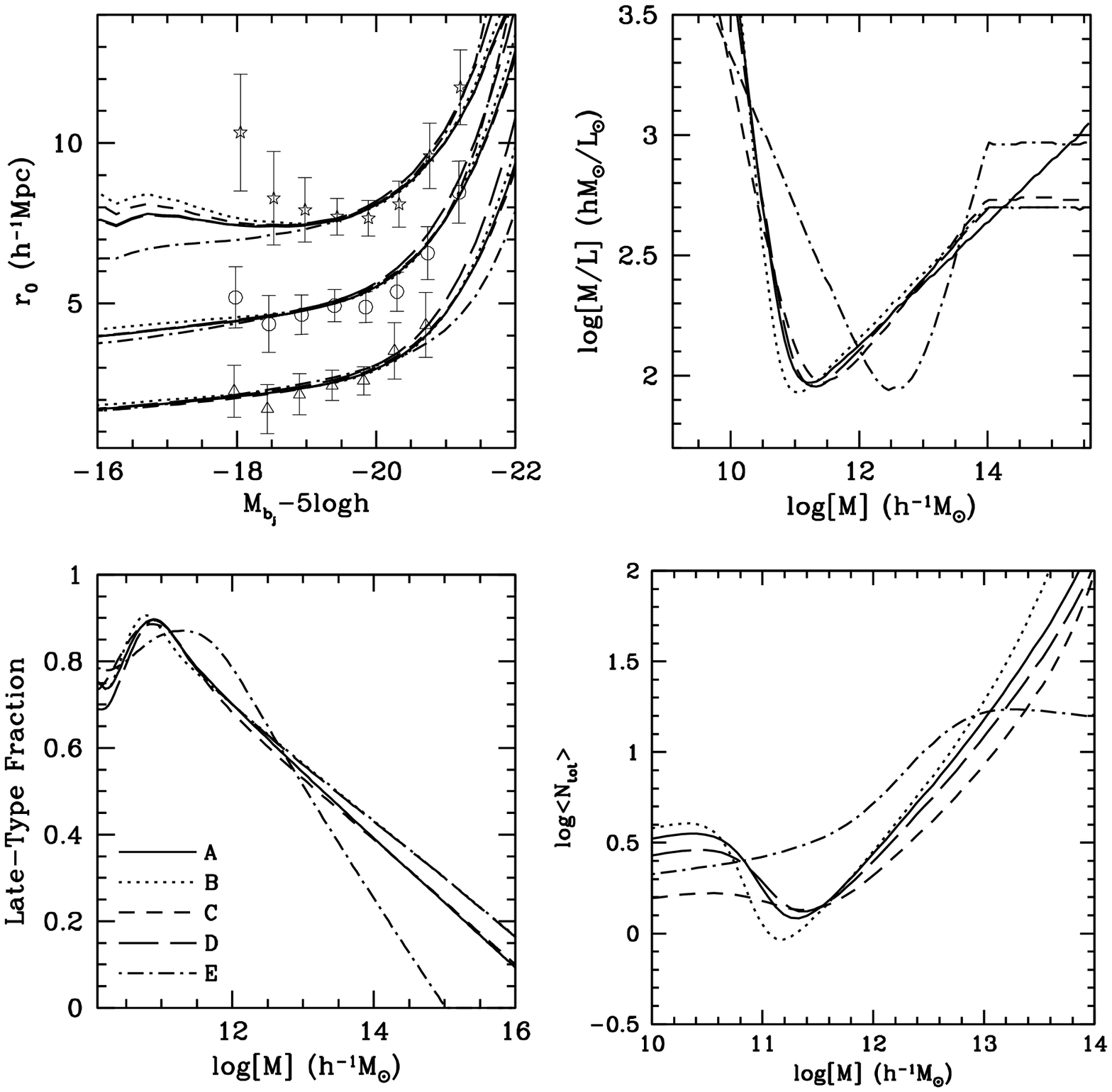,width=0.95\hdsize}}
\caption{A comparison of  the five models discussed  in the text.  The
upper panel on  the left plots  the correlation lengths as function of
luminosity. Symbols with errorbars correspond to  the 2dFGRS data with
the $r_0(L)$  of the late-type and  early-type galaxies  translated by
$-2 h^{-1} \Mpc$ and $+2 h^{-1} \Mpc$, respectively. The fits of these
models to  the observed  LFs   are not  shown, since the   differences
between the  various models  can not  be discerned by  eye.  The other
three panels plot, for   all five models, the  inferred  mass-to-light
ratios $\langle M/L \rangle(M)$ (upper right), the late-type fractions
${\cal F}_l(M)$ (lower left) and the  average total number of galaxies
$\langle  N_{\rm tot}(M) \rangle$    (lower right). Model  E  predicts
mass-to-light ratios and  late-type fractions for cluster-sized haloes
that are inconsistent with data.  In order to discriminate between the
other  models,  better  constraints on   $r_0$ for  faint,  early-type
galaxies and/or    accurate, independent  
measurements of ${\cal    F}_l(M)$ and/or
$\langle M/L \rangle$  for haloes with $M  \gta  10^{14} h^{-1} \Msun$
are required. }
\label{fig:models}
\end{figure*}

We start  our investigation using  all available model  freedom, which
means that  we minimize $\chi^2$ over  a total of  10 free parameters.
In the first step  we find the 8 parameters of the  CLF that best fits
the LF and  correlation lengths of the combined  sample. In the second
step, we then  find the $M_0$ and $M_1$,  describing $h(M)$, that best
reproduce  the  observed  clustering  properties  of  the  early-  and
late-type galaxies.   We refer to  this model with maximum  freedom as
model~A.
\begin{table*}
\begin{minipage}{\hdsize}
\caption{Model parameters.}
\begin{tabular}{lcccccccccccccc}
   \hline
ID & log$M_L$ & $(M/L)_{cl}$ & log$M_c$ & log$M_0$ & log$M_1$ & log$M_2$ &
$(M/L)_0$ & $\gamma_1$ & $\gamma_2$ & $\gamma_3$ & $\zeta$ &
$\alpha_{15}$ & $\chi^2(\Phi)$ & $\chi^2(r_0)$ \\
 (1) & (2) & (3) & (4) & (5) & (6) & (7) & (8) & (9) & (10) & 
(11) &(12) &(13) &(14) &(15) \\
\hline\hline
A  & $13.20$     & $--$      & $10.85$ & $16.64$     & $8.60$      & $12.15$ & $136$ & $2.35$  & $0.26$ & $0.73$ & $-0.19$ & $-1.09$ & $65.0$ & $5.8$ \\
B  & $13.03$     & {\bf 500} & $10.72$ & $17.27$     & $7.92$      & $12.21$ & $128$ & $3.04$  & $0.27$ & $0.73$ & $-0.19$ & $-1.20$ & $64.6$ & $5.2$ \\
C  & {\bf 13.25} & $544$     & $11.11$ & $16.71$     & $8.83$      & $11.82$ & $115$ & $1.36$  & $0.34$ & $0.65$ & $-0.39$ & $-1.27$ & $64.7$ & $5.6$ \\
D  & {\bf 13.25} & {\bf 500} & $10.94$ & $17.26$     &$10.86$      & $12.04$ & $124$ & $2.02$  & $0.30$ & $0.72$ & $-0.22$ & $-1.10$ & $64.3$ & $5.4$ \\
E  & {\bf 13.50} & $924$     & $12.67$ & $15.02$     &$10.11$      & $11.60$ &  $89$ & $0.63$  & $0.99$ & $0.69$ & $-0.20$ & $-0.73$ & $71.0$ & $6.3$ \\
B1 & $13.03$     & {\bf 500} & $10.72$ & {\bf 17.00} & {\bf 12.00} & $12.21$ & $128$ & $3.04$  & $0.27$ & $0.73$ & $-0.19$ & $-1.20$ & $64.6$ & $5.4$ \\
B2 & $13.03$     & {\bf 500} & $10.72$ & {\bf 15.00} & {\bf 13.90} & $12.21$ & $128$ & $3.04$  & $0.27$ & $0.73$ & $-0.19$ & $-1.20$ & $64.6$ & $5.6$ \\
\hline
\end{tabular}
\medskip

Column~(1) lists the  ID by which we refer to each  model in the text.
Columns~(2)  to~(13)   list  the  best-fit   model  parameters,  where
parameters  that were  kept  fixed during  the  fitting procedure  are
type-set   in   boldface.    Here   $M_L$   is   defined   such   that
$\wLstar(M_L)=L^{*}$ (see  Section~\ref{sec:res}), $(M/L)_{\rm cl}$ is
the mass-to-light  ratio of haloes  with $M \geq 10^{14}  \Msunh$, and
$\alpha_{15}$ is the faint-end slope  of the conditional LF for haloes
with $M=10^{15} h^{-1} \Msun$.   Columns~(14) and~(15) list the values
of   $\chi^2(\Phi)$   and  $\chi^2(r_0)$   of   the  best-fit   model,
respectively. Here  $\chi^2(\Phi)$ corresponds to the  $\chi^2$ of the
fit to  the LF  of the combined  sample only  ($N_{\Phi}=44$), whereas
$\chi^2(r_0)$  is summed  over  all $r_0$  measurements  of all  three
samples  (combined,  early- and  late-type;  $N_{r}=23$).  Masses  and
mass-to-light   ratios  are   in  $h^{-1}   \Msun$  and   $h  \MLsun$,
respectively.

\end{minipage}
\end{table*}

The resulting best-fit parameters are  listed in Table~1, and the fits
to the data  are shown in the upper  panels of Figure~\ref{fig:modelA}
(solid lines).  Overall  the fit to the data  is remarkably good.  The
fact that the model reproduces even the small scale features in the LF
of the early-type galaxies is a direct consequence of the way in which
we have defined $g(L)$ (see Section~\ref{sec:condLF}).

The  lower  two  panels  of Figure~\ref{fig:modelA}  plot  the  number
fractions  of late-type  galaxies  as function  of luminosity,  ${\cal
F}_l(L)$,  and  mass,  ${\cal  F}_l(M)$.  The  fraction  of  late-type
galaxies  decreases with  both increasing  luminosity and  mass.  Note
that this is, at least qualitatively, in agreement with the well-known
morphology-density relation.   The `noisy' wiggles  in ${\cal F}_l(L)$
are due to the fact that $g(L)$ is computed directly from the observed
LFs, while the feature in ${\cal F}_l(M)$ for $M \lta 3 \times 10^{10}
h^{-1} \Msun$ is  a consequence of the upturn in  the LF of early-type
galaxies at $L \lta 10^{8} h^{-2} \Lsun$.
\begin{figure}
\centerline{\psfig{figure=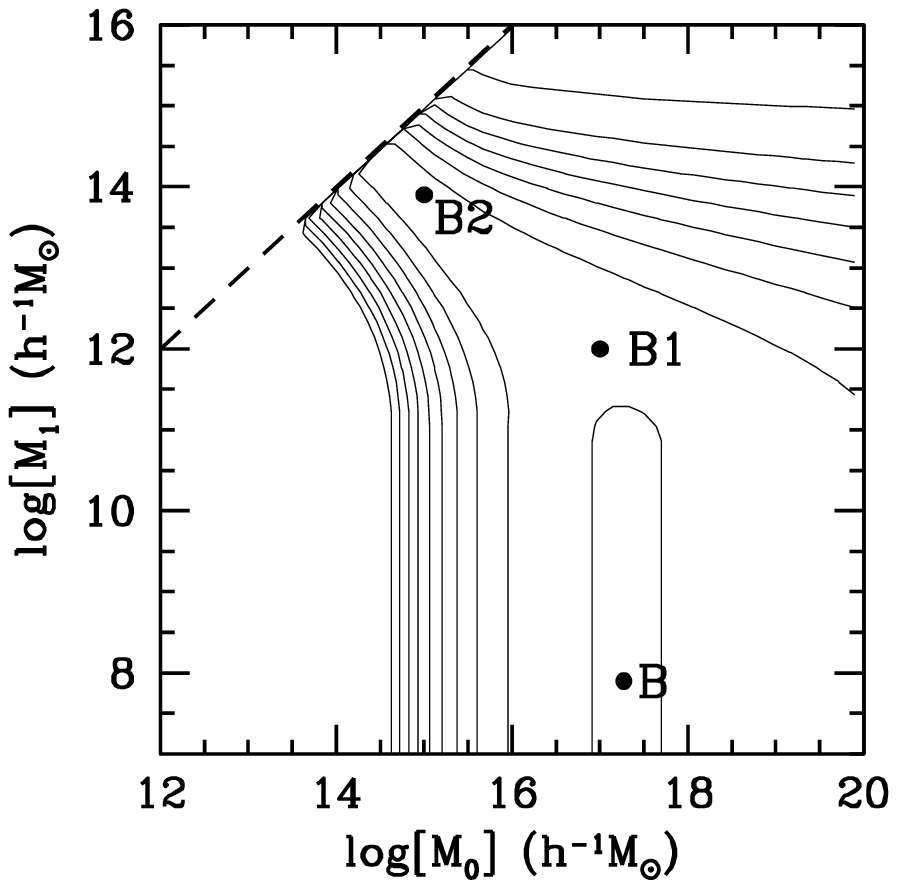,width=\hssize}}
\caption{Contour plot of $\chi^2(r_0)$ as function  of $M_0$ and $M_1$
for   model    B.       Contours  correspond  to     $\chi^2(r_0)    =
5.5,6.5,7.5,...,14.5$.   The    location  of the   best-fit   model is
indicated by a thick dot  labeled `B'. The  locations of models~B1 and
B2, discussed  in the text, are  also indicated.  The dashed, diagonal
line corresponds to $M_0 = M_1$ and indicates the boundary above which
no good fits can be obtained. For models  with $M_1 <  M_0$ there is a
large  area  in  $(M_0,M_1)$  parameter space  that  yields  virtually
identical $\chi^2(r_0)$.}
\label{fig:chi2}
\end{figure}

The  upper-right  panel of  Figure~\ref{fig:models}  plots the  total,
average  mass-to-light ratio  of  model~A as  function  of halo  mass,
computed  using equation~(\ref{LofM}).   The $\langle  M/L \rangle(M)$
reveals a sharp  minimum of $\sim 110 h \MLsun$ at  $M \simeq 2 \times
10^{11} h^{-1}  \Msun$.  For $M <  2 \times 10^{11}  h^{-1} \Msun$ the
mass-to-light ratios increase  dramatically with decreasing mass, such
that in haloes with $M  \lta 10^{10} h^{-1} \Msun$ virtually no (blue)
light is produced.  This sharp  increase of $\langle M/L \rangle$ with
decreasing halo mass is required in  order to bring the steep slope of
the halo  mass function  at low $M$  in agreement with  the relatively
shallow faint-end slope of the observed LF.  In the language of galaxy
formation models;  low-mass haloes need efficient  feedback to prevent
an overabundance  of faint  galaxies.  For haloes  with $M >  2 \times
10^{11}  h^{-1} \Msun  $ the  (average) mass-to-light  ratio increases
roughly as  $\langle M/L \rangle \propto M^{0.3}$,  such that clusters
of  $10^{15} h^{-1} \Msun$  have, on  average, a  (blue) mass-to-light
ratio of $\sim 1000 h \MLsun$.  Within the context of galaxy formation
models,  this increase  in  mass-to-light ratio  is  interpreted as  a
decrease of the cooling efficiency in more massive haloes (e.g., White
\& Rees 1978; White \& Frenk 1991).

By construction, $\langle M/L \rangle \propto M^{\gamma_2}$ for haloes
with $M  \gg M_c$.   However, several studies  have suggested  that on
cluster mass scales $\langle M/L \rangle$ varies only weakly with mass
(e.g.,  Bahcall, Lubin \&  Norman 1995;  Bahcall \etal  2000; Kochanek
\etal 2002).  Furthermore, Fukugita,  Hogan \& Peebles (1998), using a
variety  of  observational   constraints,  derived  that  clusters  of
galaxies have  blue mass-to-light  ratios of $450  \pm 100  h \MLsun$,
i.e., more than $5 \sigma$ lower than what model A predicts for a Coma
sized cluster.  Therefore, we now  consider a model (model~B) in which
we keep $\langle M/L \rangle$ fixed at a constant value of $(M/L)_{\rm
cl}$  for haloes  with $M  \geq  10^{14} h^{-1}  \Msun$. Motivated  by
Fukugita  \etal (1998)  we  adopt  $(M/L)_{\rm cl}  =  500 h  \MLsun$.
Continuity  of  $\langle  M/L  \rangle(M)$  then  sets  the  parameter
$\gamma_2$  which  is therefore  no  longer  a  free parameter.   With
respect to model~A, the minimum  of $\langle M/L \rangle(M)$ occurs at
a somewhat lower mass of  $\sim 10^{11} h^{-1} \Msun$, the fraction of
late-type galaxies in massive haloes has increased, and the fit to the
correlation lengths has improved,  especially for the faint early-type
galaxies (see  Figure~\ref{fig:models}).  This is  easy to understand:
Lowering the  mass-to-light ratios on  cluster scales means  that more
galaxies have  to reside in  clusters.  Since more massive  haloes are
more strongly clustered (Mo \& White 1996) and the majority of cluster
galaxies are early-types, this leads  to an increase of $r_0$ which is
most pronounced for the early-types.

Although overall  model~B is  in better agreement  with the  data than
model~A, it also  has an unattractive feature. This  is illustrated in
the  lower right panel  of Figure~\ref{fig:models}  where we  plot the
average, total number of galaxies
\begin{equation}
\label{ntotal}
\langle N_{\rm tot}(M) \rangle = \int_{0}^{\infty} \Phi(L \vert M) \,
{\rm d}L = \tilde{\Phi}^{*} \, \Gamma(\walpha+1)
\end{equation}
as function  of halo mass. Model~B  predicts that this  number is less
than unity for haloes with  masses around $10^{11} h^{-1} \Msun$. Yet,
this  mass scale  coincides with  the  {\it minimum}  of $\langle  M/L
\rangle(M)$.  This  implies that these  haloes must have  an extremely
large spread  in $M/L$; the  majority of haloes contain  zero galaxies
(or only  ``dark'' galaxies,  which produce no  light), while  a small
fraction harbors one  (or more) relatively bright galaxy  (in order to
explain the $\langle  M/L \rangle$).  This aspect of  the model is due
to  the  fact that  the  CLF  for haloes  of  this  mass  scale has  a
relatively  high characteristic  luminosity $\wLstar$.   If  we define
$M_L$ as the  mass for which $\wLstar(M) = L^{*}$,  where $L^{*} = 1.1
\times 10^{10}  h^{-2} \Lsun$ is the characteristic  luminosity of the
{\it observed} LF of the combined sample (see Madgwick \etal 2002), we
find for model~B a relatively  low value of $M_L \simeq 10^{13} h^{-1}
\Msun$ (see Table~1).  Since the abundance of haloes with this mass is
relatively  high,  reproducing   the  observed  abundance  of  $L^{*}$
galaxies requires  a significant  fraction of haloes  with dark  or no
galaxies.  Although  the existence of dark galaxies  is interesting in
itself (see Verde,  Oh \& Jimenez 2002), a large  scatter in $M/L$ for
haloes with  $M \simeq 10^{11}  h^{-1} \Msun$ seems  inconsistent with
the small  scatter in the observed Tully-Fisher  and Fundamental Plane
relations (see also discussion in Paper~1).
\begin{figure*}
\centerline{\psfig{figure=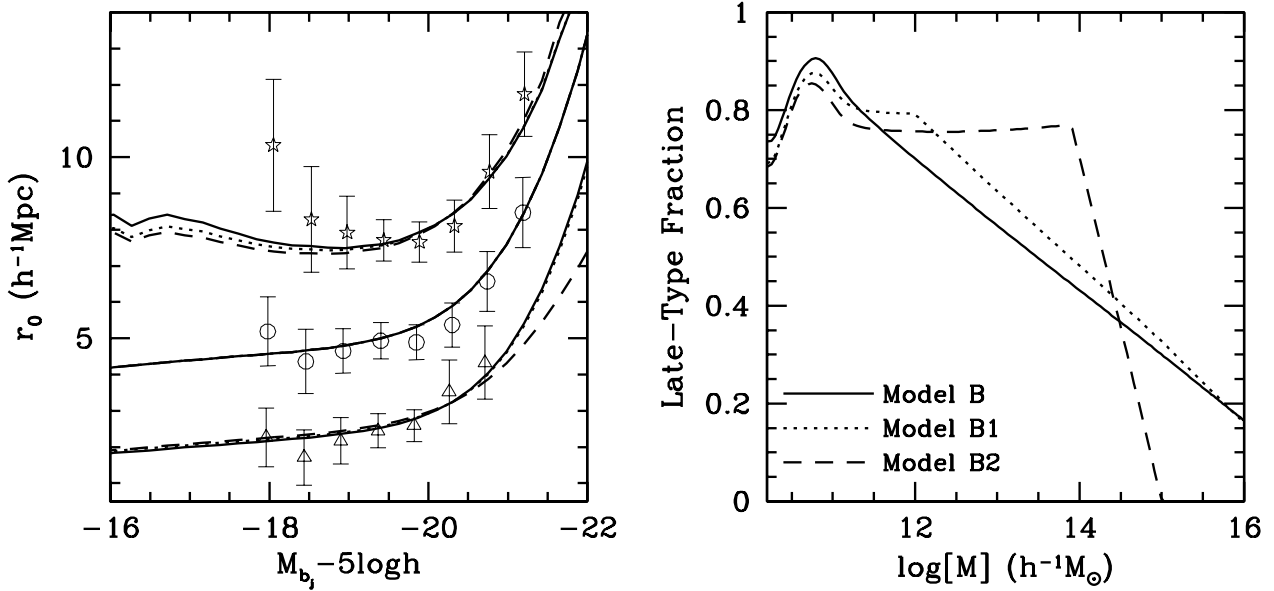,width=0.95\hdsize}}
\caption{Correlation  lengths and late-type  fractions as  function of
magnitude  and halo  mass,  respectively, for  models~B,  B1, and  B2.
Although   these  three   models  yield   virtually  indistinguishable
$r_0(L)$, they  imply ${\cal F}_l(M)$ that are  quite different.  Note
that  model~B2  can  be  rejected  based on  its  prediction  of  zero
late-type galaxies in haloes with $M > 10^{15} h^{-1} \Msun$.}
\label{fig:degen}
\end{figure*}

Therefore, in models~C and~E we tune the value of $M_2$ such that $M_L
= 10^{13.25} h^{-1} \Msun$  and $10^{13.5} h^{-1} \Msun$, respectively
(i.e.,  $M_2$ is  not  a free  parameter  here).  We  again set  $M/L$
constant for haloes with $M \geq 10^{14} h^{-1} \Msun$, but we let the
value of $(M/L)_{\rm cl}$ be a free parameter.  These models thus have
the same  number of free parameters  as model~B.  As can  be seen from
Figure~\ref{fig:models},    model~C   is    remarkably    similar   to
model~A. This  is mainly due  to the fact  that the value of  $M_L$ of
model~A is with  $10^{13.26}$ close to that of  model~C (see Table~1).
The  best-fit  value  of  $(M/L)_{\rm  cl}$ is  with  $615  h  \MLsun$
consistent with the value advocated by Fukugita \etal (1998) at the $2
\sigma$  level.   Model~E, however,  is  dramatically different.   The
minimum of  $\langle M/L \rangle(M)$ occurs at  a significantly higher
mass scale  of $\sim 2.5  \times 10^{12} h^{-1} \Msun$.   In addition,
$(M/L)_{\rm cl} \simeq 1000 h \MLsun$  and ${\cal F}_l = 0$ for haloes
with $M  \gta 10^{15}  h^{-1} \Msun$, both  of which  are inconsistent
with observations. Furthermore,  model~E yields a significantly poorer
fit  to  the observed  correlation  lengths  of  the faint  early-type
galaxies, and we therefore no longer consider model~E in what follows.

Finally, in  model~D, we combine  the constraints from  models~B and~C
above; model~D has therefore only 8 free parameters together with $M_L
= 10^{13.25} h^{-1}  \Msun$ and $(M/L)_{\rm cl} =  500 h \MLsun$. With
respect to model~C this causes a small increase in correlation lengths
(again because more  light is added to the  strongly clustered massive
haloes),  though the  effect is  sufficiently small  that  model~D can
still be considered as yielding an overall good fit.

Figure~\ref{fig:chi2} plots contours  of constant $\chi^2(r_0)$ in the
$M_0$ versus  $M_1$ plane  for model~B (results  for the  other models
discussed  above   are  very  similar).   The   dashed  diagonal  line
corresponds  to  $M_0  =  M_1$  (i.e.,  $h(M)$  is  a  step-function).
Clearly, models with $M_0 <  M_1$ (for which the fraction of late-type
galaxies {\it  increases} with halo mass)  result in poor  fits as the
value of  $\chi^2(r_0)$ is  always extremely large.   This is  easy to
understand.  Since  the correlation lengths of  the late-type galaxies
are  smaller than  those of  the early-type  galaxies, and  since more
massive haloes  are more strongly clustered,  reproducing the observed
$r_0(L)$  requires that  the  late-type fraction  decreases with  halo
mass, and  thus that $M_0 >  M_1$.  However, as is  also apparent from
the contour plot, a large area in the $M_0$-$M_1$ plane yields roughly
equally  good fits.  To  illustrate this  ``degeneracy'' we  have also
computed  two additional  models, B1  and B2,  for which  we  keep all
parameters identical to that of model~B, but we modify $M_0$ and $M_1$
so  that  these  models  fall  in  the  region  of  roughly  the  same
$\chi^2(r_0)$  (indicated  by  solid dots  in  Figure~\ref{fig:chi2}).
Note that changes in $h(M)$ only affect the $r_0(L)$ of the early- and
late-type  galaxies;  all  LFs  and  the correlation  lengths  of  the
combined sample are unaffected.   The results for models~B1 and~B2 are
shown  in  Figure~\ref{fig:degen}.  As  is  also  apparent from  their
values of $\chi^2(r_0)$ listed  in Table~1, these models are virtually
indistinguishable from model B and  from each other.  However, they do
yield  somewhat  different ${\cal  F}_l(M)$.   Model  B2 for  instance
yields late-type fractions of zero  for $M \gta 10^{15} h^{-1} \Msun$.
This is inconsistent  with data, which allows us  to rule against this
particular model.  However, models B1  and B have quite similar ${\cal
F}_l(M)$, and  current data is not sufficient  to discriminate between
these  two alternatives.  This  degeneracy also  explains why  we have
adopted a fairly simple form  for $h(M)$; more complicated forms, with
more free parameters, only increases the amount of degeneracy.

Having shown that there are different models that fit the data roughly
equally well, the question arises whether one can discriminate between
these  different  halo  occupancy   models.   Based  on  the  observed
mass-to-light  ratios  of  clusters,  and  on the  amount  of  scatter
inferred from  the Tully-Fisher  and Fundamental Plane  relations, one
might  argue that  of the  models presented  above, model~C  is  to be
preferred  (modulo some uncertainties  in $M_0$  and $M_1$).   In what
follows, however, we always compare predictions from different models.
This  gives an  idea  about the  extent  of the  uncertainties in  the
models, and the accuracy in the data required to allow to discriminate
between  the various  models.  Some  clues are  already  apparent from
Figure~\ref{fig:models}.   As  we have  shown  above,  the models  are
extremely  sensitive to  the  exact value  of  $M_L$ and  most of  the
uncertainties in  the models can be  translated to values  of $M_L$ in
the range $10^{13} h^{-1} \Msun$ to $10^{13.5} h^{-1} \Msun$. In order
to further constrain this crucial parameter more accurate measurements
are required  of the mass-to-light  ratios and late-type  fractions in
clusters or of  the correlation lengths of galaxies  with $M_{b_J} - 5
{\rm log}h \gta -18$.

In  summary, within the  concordance cosmology  several models  can be
found  that fit  the data  remarkably  well. Although  some amount  of
degeneracy  exists, the  mean trends  reveal an  average mass-to-light
which has  a minimum  of $\sim 100  \, h  \MLsun$ at around  $3 \times
10^{11} h^{-1}  \Msun$, and which increases  extremely rapidly towards
lower masses:  in all cases we find  $\langle M/L \rangle >  3000 \, h
\MLsun$ for haloes  with $M < 10^{10} h^{-1}  \Msun$.  The correlation
lengths require that the fraction of late-type galaxies decreases from
about  90 percent for  haloes with  $M \lta  10^{12} h^{-1}  \Msun$ to
anywhere  between 0  and 40  percent  on scales  of $M=10^{15}  h^{-1}
\Msun$, in qualitative agreement with the morphology-density relation.
Similarly, the observed LFs  indicate a similar trend with luminosity:
about 90 percent of all galaxies with $L \lta 10^{9} h^{-2} \Lsun$ are
late-types,  while this  drops  to about  20-30  percent for  galaxies
brighter than $L^{*}$.

\section{Two Point Correlation Functions}
\label{sec:corrfunc}

Two  point correlation functions  are a powerful  tool to describe the
clustering  properties  of  galaxies.  With the   CLFs  for early- and
late-type  galaxies  we  can  compute  the   galaxy-galaxy correlation
function, $\xi_{\rm gg}(r)$, not only as function  of galaxy type, but
also  as function of luminosity.  This allows a detailed investigation
of the  bias  of galaxies  as  function   of scale,  luminosity,   and
(spectral) type.

At small scales, where the number  of pairs are mostly due to galaxies
within  the same  halo,  $\xi_{\rm  gg}(r)$ depends  not  only on  the
occupation numbers of  galaxies (which can be obtained  from the CLF),
but also on how galaxies are distributed inside individual dark matter
haloes.   In  addition,  since  the  number of  pairs  within  a  halo
containing $N$  galaxies is equal  to ${1 \over  2} N (N-1)$  one also
needs  to  know  the  second  moment of  the  halo  occupation  number
distribution.  Since the CLF only contains information about the first
moments $\langle N(M) \rangle$ we need to make additional assumptions.
Alternatively, as  shown by  Cooray (2002), one  may use  the observed
power spectrum of galaxies to obtain constraints on the second moments
of $P(N \vert M)$ using standard inversion techniques.
\begin{figure*}
\centerline{\psfig{figure=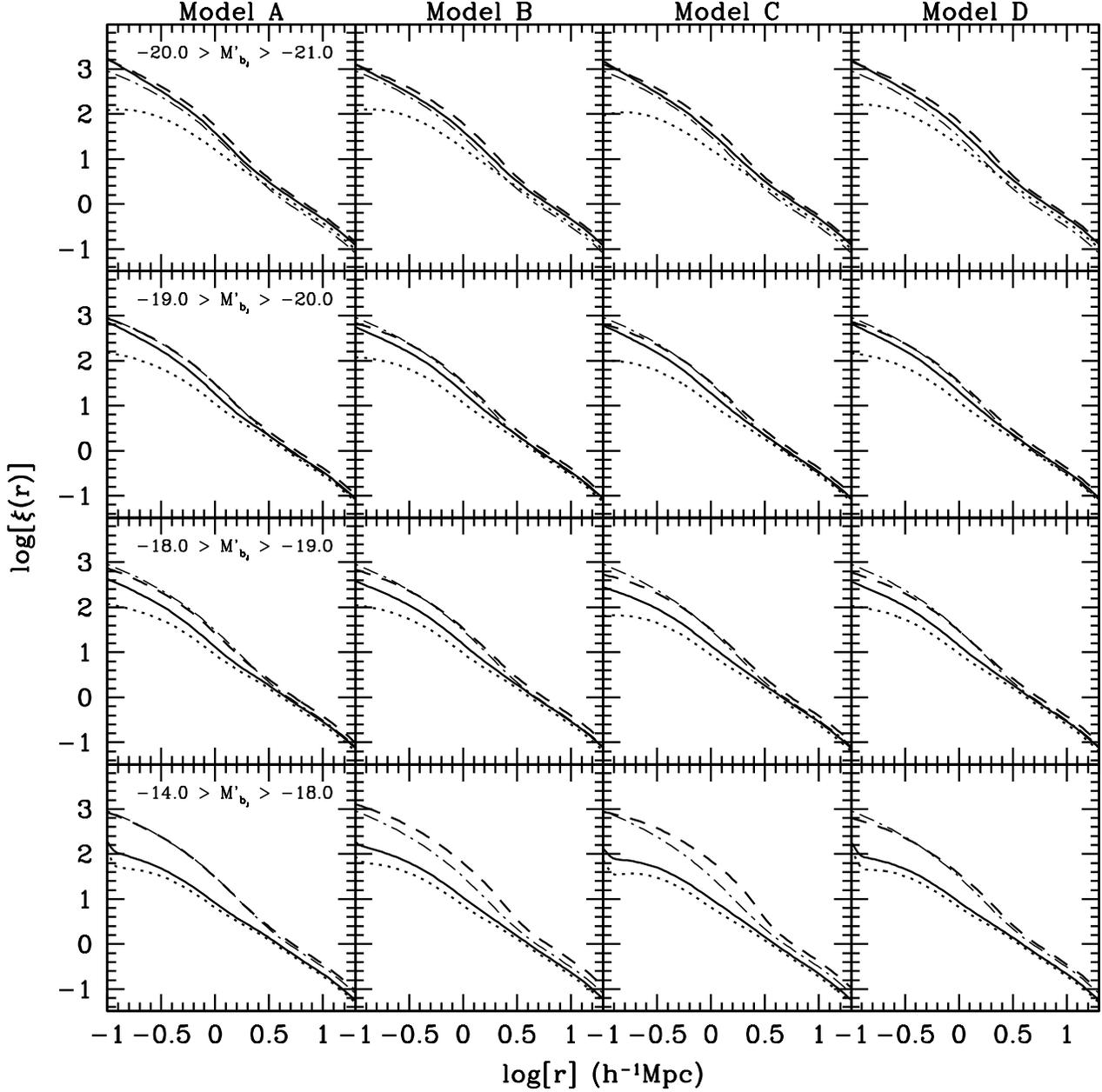,width=0.95\hdsize}}
\caption{Real-space  correlation  functions  of  all  galaxies  (solid
lines),  late-type galaxies  (dotted lines),  and  early-type galaxies
(dashed lines). Results  are shown for four different  models, and for
four different magnitude bins (as indicated).  Note that $M'_{b_J}$ is
defined as  $M_{b_J}- 5  {\rm log}h$.  The  dot-dashed curves  in each
panel  correspond  to  the  real-space  correlation  function  of  the
evolved, non-linear dark matter  mass distribution, and is plotted for
comparison.  Note that the  ratio between the correlation functions of
early- and late-type galaxies increases with decreasing $r$.  See text
for a detailed discussion.}
\label{fig:xi}
\end{figure*}

We  follow  Yang  \etal  (2002)  and  make  the  assumption  that  the
probability distribution $P(N \vert M)$, with $N$ an integer, is given
by
\begin{equation}
\label{pnm}
P(N \vert M) = \left\{ \begin{array}{lll}
N' + 1 - \langle N(M) \rangle & \mbox{if $N = N'$} \\
\langle N(M) \rangle - N'     & \mbox{if $N = N' + 1$} \\
0 & \mbox{otherwise}
\end{array} \right.
\end{equation}
Here $N'$ is the integer of $\langle N(M) \rangle$.  Thus, the actual,
integer number  of galaxies in  a halo of  mass $M$ is either  $N'$ or
$N'+1$.  This particular model for the distribution of halo occupation
numbers  is supported by  the semi-analytical  models of  Benson \etal
(2000) who found that the  halo occupation probability distribution is
narrower than  a Poisson distribution.   In addition, they  have shown
that  distribution~(\ref{pnm}),   which  they  call   the  ``average''
distribution,   is  successful   in  yielding   power-law  correlation
functions, much more  so than for example a  Poisson distribution (see
also Berlind \& Weinberg  2002). For this ``average'' distribution the
average number of  galaxy pairs inside an individual  dark matter halo
is given by
\begin{equation}
\label{Npair}
\langle N_{\rm pair}(M) \rangle =  N' \, \langle N(M) \rangle - 
{1 \over 2} N' \, (N' + 1)
\end{equation}
\begin{figure*}
\centerline{\psfig{figure=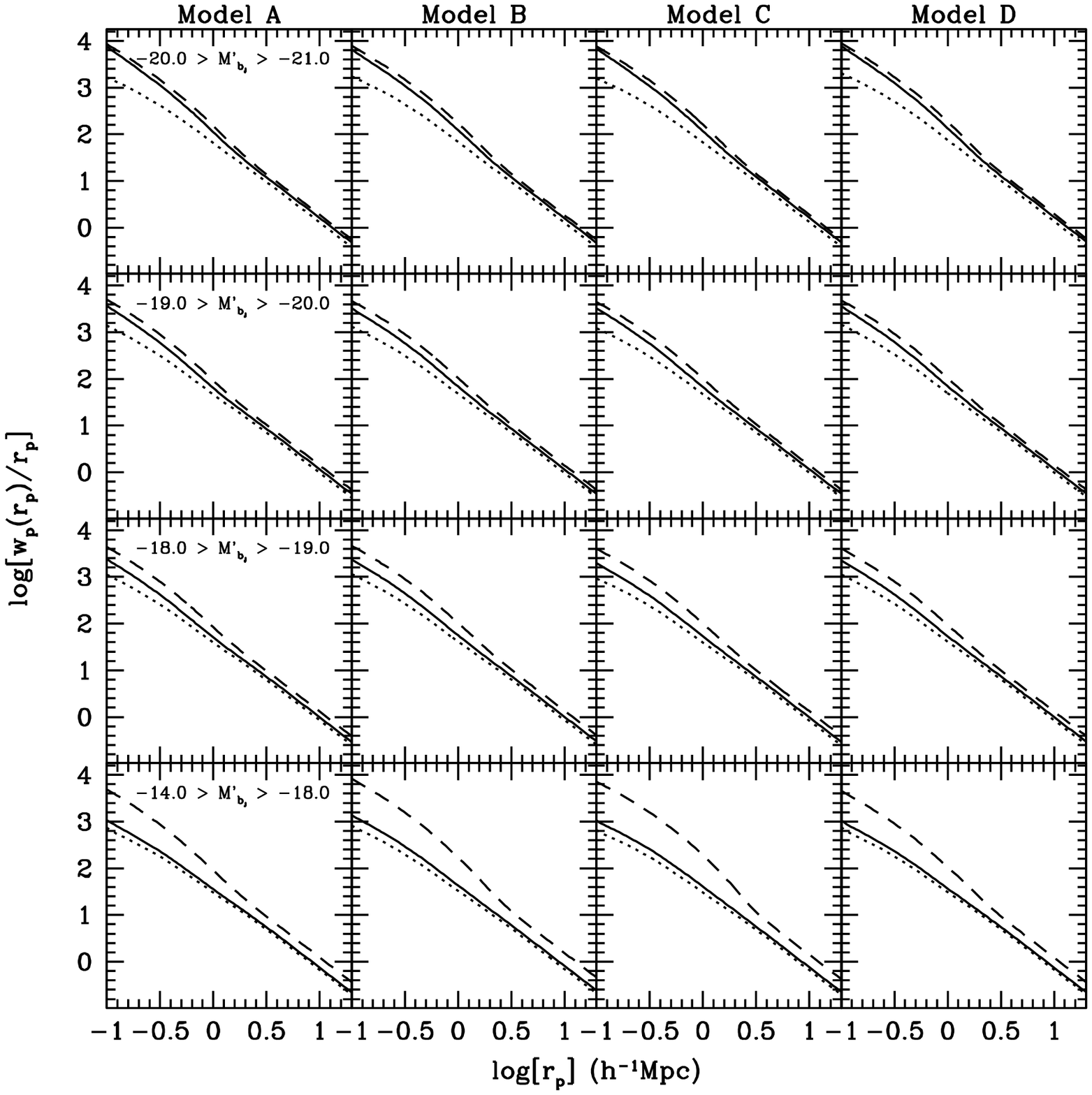,width=0.95\hdsize}}
\caption{Same  as Figure~\ref{fig:xi},  except that  here we  plot the
projected   correlation  function   $w_p(r_p)/r_p$   instead  of   the
real-space correlation  function $\xi(r)$.  Note  that these projected
correlation  functions,   which  unlike  the   real-space  correlation
functions  can be directly  obtained from  redshift surveys,  are much
smoother  than the  real-space correlation  functions and  much better
resemble single power-laws.}
\label{fig:xiproj}
\end{figure*}

For  the spatial  distribution of  galaxies  within a  halo we  follow
Peacock \& Smith  (2000), Benson \etal (2000) and  Berlind \& Weinberg
(2002) and assume that (i) the brightest galaxy resides in the center,
and  (ii)  the  remaining  `satellite'  galaxies  follow  the  density
distribution of the dark matter  (which we specify in Appendix~A). The
special treatment of a central galaxy is required if the galaxy-galaxy
correlation function is to remain  close to a single power-law, rather
than  to  reveal a  flattening  as present  in  the  dark matter  mass
correlation  function  (Peacock \&  Smith  2000;  Berlind \&  Weinberg
2002). Pairs between the  central galaxy and satellite galaxies follow
a separation function $f_{\rm cs}(r)  = 4 \pi \tilde{\rho}(r) \, r^2$,
with  $\tilde{\rho}(r)$  the normalized  density  distribution of  the
halo,   which   integrates   to   a   total  halo   mass   of   unity.
Satellite-satellite  pairs  follow  a  different  separation  function
$f_{\rm  ss}(r)$ which  can also  be  obtained from  the halo  density
distribution. The  separation function for  all pairs, $f(r)$,  can be
written as
\begin{eqnarray}
\label{Npairrad}
\langle N_{\rm pair}(M) \rangle f(r) \, {\rm d}r & = & 
\langle N_{\rm cs}(M) \rangle \, f_{\rm cs}(r) \, {\rm d}r + \nonumber \\
 & &  \langle N_{\rm ss}(M) \rangle \, f_{\rm ss}(r) \, {\rm d}r
\end{eqnarray}
where  $\langle  N_{\rm cs}(M)  \rangle$  and  $\langle N_{\rm  ss}(M)
\rangle$  correspond to  the average  number of  central-satellite and
satellite-satellite pairs, respectively.

Once the CLF and $\langle N_{\rm pair}(M) \rangle f(r)$ are specified,
$\xi_{\rm    gg}(r)$    can     be    computed    (see    Appendix~A).
Figure~\ref{fig:xi} plots $\xi_{\rm gg}(r)$ for the late-type galaxies
(dotted  lines),  the  early-type  galaxies  (dashed  lines)  and  the
combined  sample of  galaxies (solid  lines) for  different luminosity
bins  and  models.   In  addition,   we  plot  the  dark  matter  mass
correlation function (dot-dashed lines)  for comparison. Panels in the
upper three rows show results  for galaxies in magnitude intervals for
which data from the 2dFGRS  exists (Norberg \etal 2002).  The panel in
the  lower row  corresponds  to  $-14 >  M_{b_J}  - 5  {\rm  log} h  >
-18$.  Since Norberg  \etal only  presented correlation  functions for
galaxies  with $M_{b_J}  -  5 {\rm  log}  h \leq  -17.5$,  no data  on
$\xi_{\rm gg}(r)$ exists for this magnitude range.
\begin{figure*}
\centerline{\psfig{figure=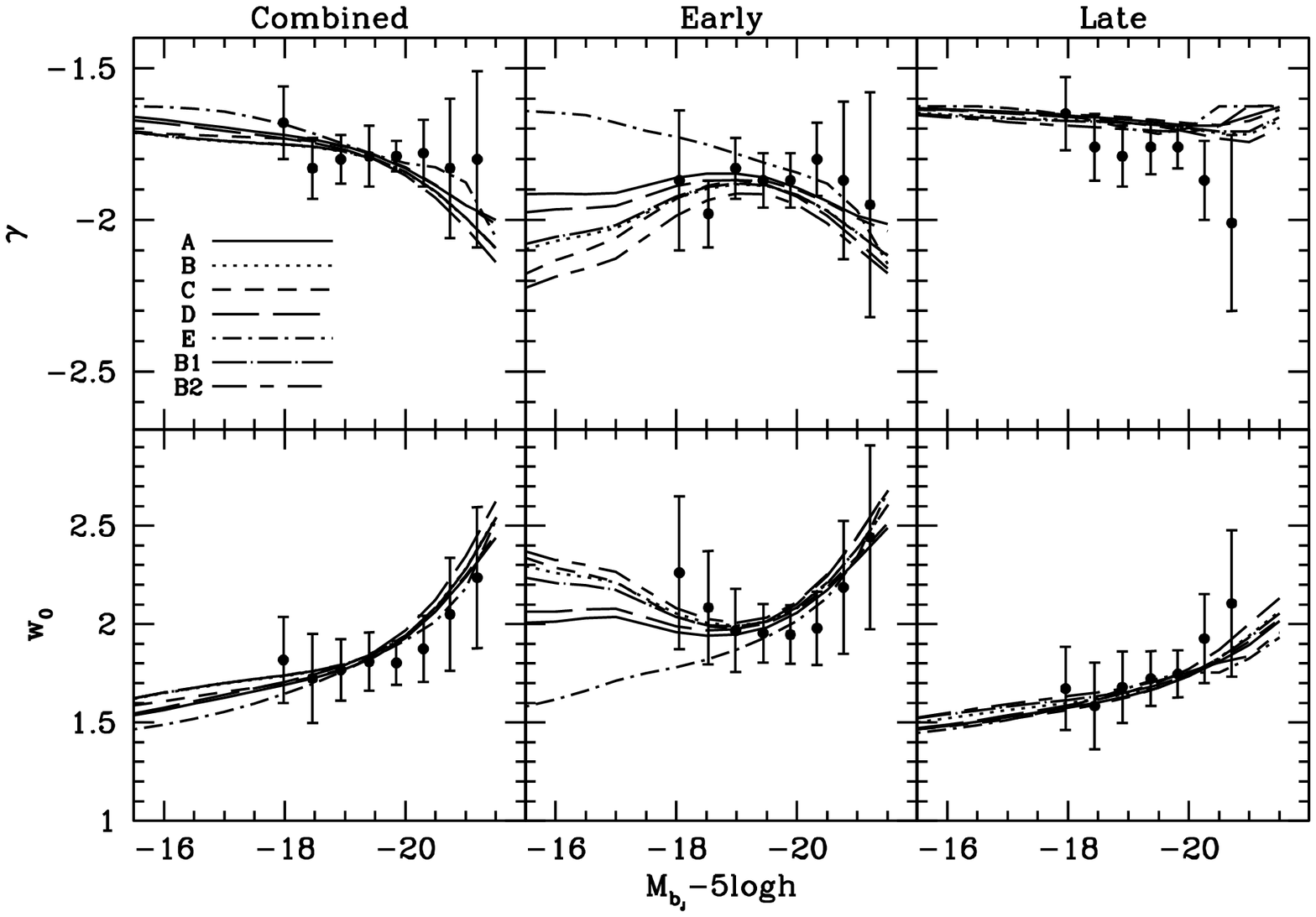,width=0.95\hdsize}}
\caption{The power-law  slopes $\gamma$  and zero-points $w_0$  of the
projected  correlation functions  of galaxies  within a  one magnitude
luminosity bin,  obtained from a  single power-law fit over  the range
$-0.5  \leq   {\rm  log}(r_p)  \leq  1.3$.    Symbols  with  errorbars
correspond  to the  2dFGRS data  of  Norberg \etal  (2002), while  the
various lines indicate predictions from our models, as indicated.  The
overall  agreement  with  the  data  is reasonable.   Except  for  the
relatively faint early-type  galaxies, differences between the various
models  are small.   Further  constraints will  have  to await  larger
redshift surveys (i.e., smaller errorbars).}
\label{fig:gamma}
\end{figure*}

Virtually  all  correlation  functions,  including that  of  the  dark
matter, reveal a special feature in the radial interval $1 h^{-1} \Mpc
\lta r \lta  3 h^{-1} \Mpc$.  This coincides  with the transition from
the 1-halo  term to  the 2-halo  term (see Appendix~A):  at $r  \lta 1
h^{-1} \Mpc$  the correlation functions  are governed by  galaxy pairs
within  individual dark  matter  haloes and  therefore  depend on  our
assumptions regarding $\langle N_{\rm  pair}(M) \rangle f(r)$.  For $r
\gta  3 h^{-1} \Mpc$,  however, the  galaxy correlation  functions are
independent  of how galaxies  are distributed  inside haloes  and only
depend  on  the first  moments  $\langle  N(M)  \rangle$ of  the  halo
occupation number distributions.

Typically,  at $r \lta  2 h^{-1}  \Mpc$ the  ratio of  the correlation
functions of  the early- and late-type  galaxies is larger  than at $r
\gta 2  h^{-1} \Mpc$.  Since  at small scales the  clustering strength
depends on the second moment of the occupancy numbers, the correlation
function  at small  $r$ is  more  dominated by  the contribution  from
massive haloes  than at  large $r$.  The  ratio between  the $\xi_{\rm
gg}(r)$ of  early- and  late-type galaxies at  small $r$  is therefore
sensitive  to  ${\cal F}_l(M)$  at  large  $M$.   This is  immediately
apparent   from  a   comparison   with  the   lower   left  panel   of
Figure~\ref{fig:models},  which  shows that  models~A  and~C are  very
similar   with  lower   late-type  fractions   than   models~B  and~D.
Consequently, the  ratio of $\xi_{\rm gg}(r)$ of  early- and late-type
galaxies at small  $r$ is larger for models~A  and~C than for models~B
and~D.
\begin{figure*}
\centerline{\psfig{figure=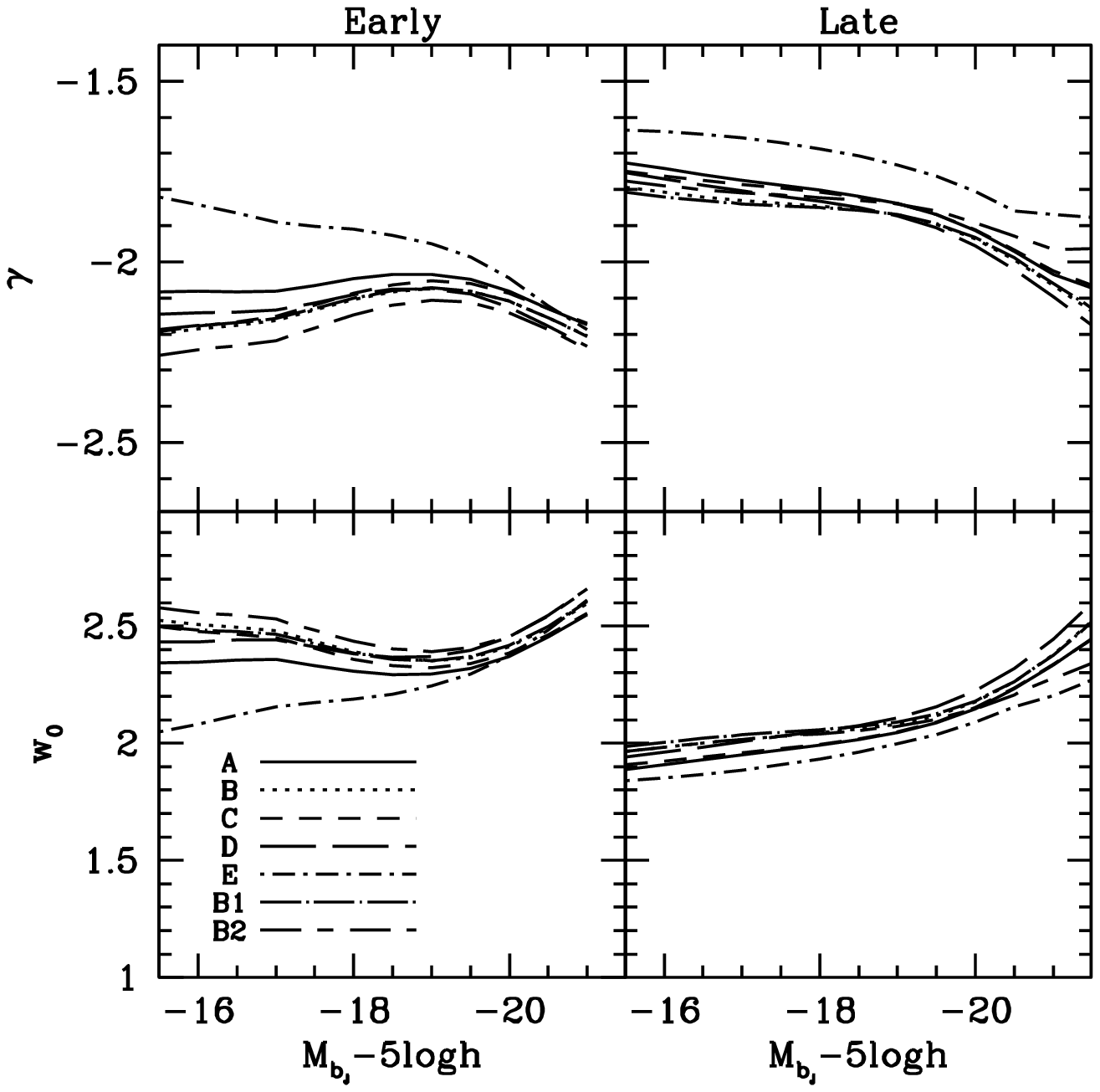,width=0.7\hdsize}}
\caption{The power-law  slopes $\gamma$  and zero-points $w_0$  of the
projected cross-correlation functions with  bright ($-21 > M_{b_j} - 5
{\rm log} h >  -22$) early-type galaxies. As in Figure~\ref{fig:gamma}
$\gamma$ and $w_0$  are obtained from a single  power-law fit over the
range $-0.5 \leq {\rm log}(r_p) \leq 1.3$.}
\label{fig:cross}
\end{figure*}

As is also immediately  apparent from Figure~\ref{fig:xi}, the bias of
galaxies  with respect to  the dark  matter mass  distribution depends
sensitively on galaxy  luminosity, on the radial scale,  and on galaxy
type  (see also  Kauffmann,  Nusser \&  Steinmetz  1997). In  general,
early-type galaxies are biased tracers of the mass distribution at all
scales,  and  for all  luminosities,  whereas  late-type galaxies  are
virtually  always  anti-biased.  The  combined  sample  of early-  and
late-type galaxies  reveals a transition  from positive bias  at large
scales ($r \gta 2 h^{-1}\Mpc$)  to anti-bias at smaller scales, except
for the most luminous galaxies which are always positively biased.

\subsection{Comparison with observations}
\label{sec:compobs}

Can  we compare  these $\xi_{\rm  gg}(r)$ to  those obtained  from the
2dFGRS? Norberg  \etal (2002)  argued that the  real-space correlation
functions  of all  subsamples of  the 2dFGRS  data are  well fit  by a
single power-law  (independent of luminosity or  spectral type). Taken
at face value, the fact that virtually all correlation functions shown
in  Figure~\ref{fig:xi} reveal clear  deviations from  pure power-laws
would force  us to rule  against each of  our models.  However,  it is
important to  take into  account how real-space  correlation functions
are  obtained from  data.   After all,  the  data only  yields a  {\it
redshift}-space correlation function,  which is distorted with respect
to the real-space correlation function  due to the peculiar motions of
the  galaxies.  On  small  scales the  virialized  motion of  galaxies
within dark matter  haloes cause a reduction of  the correlation power
(the so-called  ``finger-of-God'' effect), while on  larger scales the
correlations are boosted  due to coherent flows (Kaiser  1987).  It is
common practice  to therefore compute the  galaxy correlation function
on  a two-dimensional grid  of pair  separations parallel  ($\pi$) and
perpendicular  ($r_p$)  to  the  line-of-sight.  Integration  of  this
$\xi_{\rm gg}(r_{p},\pi)$  over $\pi$ then  yields what is  called the
projected  correlation function,  which is  related to  the real-space
correlation function by an Abel transform
\begin{equation}
\label{abel}
w_p(r_p) = \int_{-\infty}^{\infty} \xi_{\rm gg}(r_p,\pi) {\rm d}\pi 
= 2 \int_{r_p}^{\infty} \xi_{\rm gg}(r) \, 
{r \, {\rm d}r \over \sqrt{r^2 - r_p^2}}
\end{equation}
(Davis \& Peebles  1983). If the real-space correlation  function is a
single  power-law,  $\xi_{\rm  gg}(r)  =  (r/r_0)^{\gamma}$  then  the
projected  correlation  function   also  follows  a  single  power-law
$w_p(r_p)/r_p = A(\gamma) (r_p/r_0)^{\gamma}$ with
\begin{equation}
\label{pxi}
A(\gamma)={\Gamma(1/2) \, \Gamma(-(\gamma+1)/2)\over\Gamma(-\gamma/2)}
\end{equation}
(see e.g., Baugh 1996).

In  Figure~\ref{fig:xiproj}  we  plot $w_p(r_p)/r_p$  (obtained  using
equation~[\ref{abel}])  for  the  same   samples  of  galaxies  as  in
Figure~\ref{fig:xi}.   As  can be  seen,  the  projection has  largely
washed-out  the features  at $\sim  2 h^{-1}  \Mpc$ in  the real-space
correlation functions, and overall the projected correlation functions
better resemble single power-laws.   Only the correlation functions of
early-type galaxies  with $-14 > M_{b_J} -  5 {\rm log} h  > -18$, for
which  no  data  exists  to  date,  clearly  deviates  from  a  single
power-law.
\begin{figure*}
\centerline{\psfig{figure=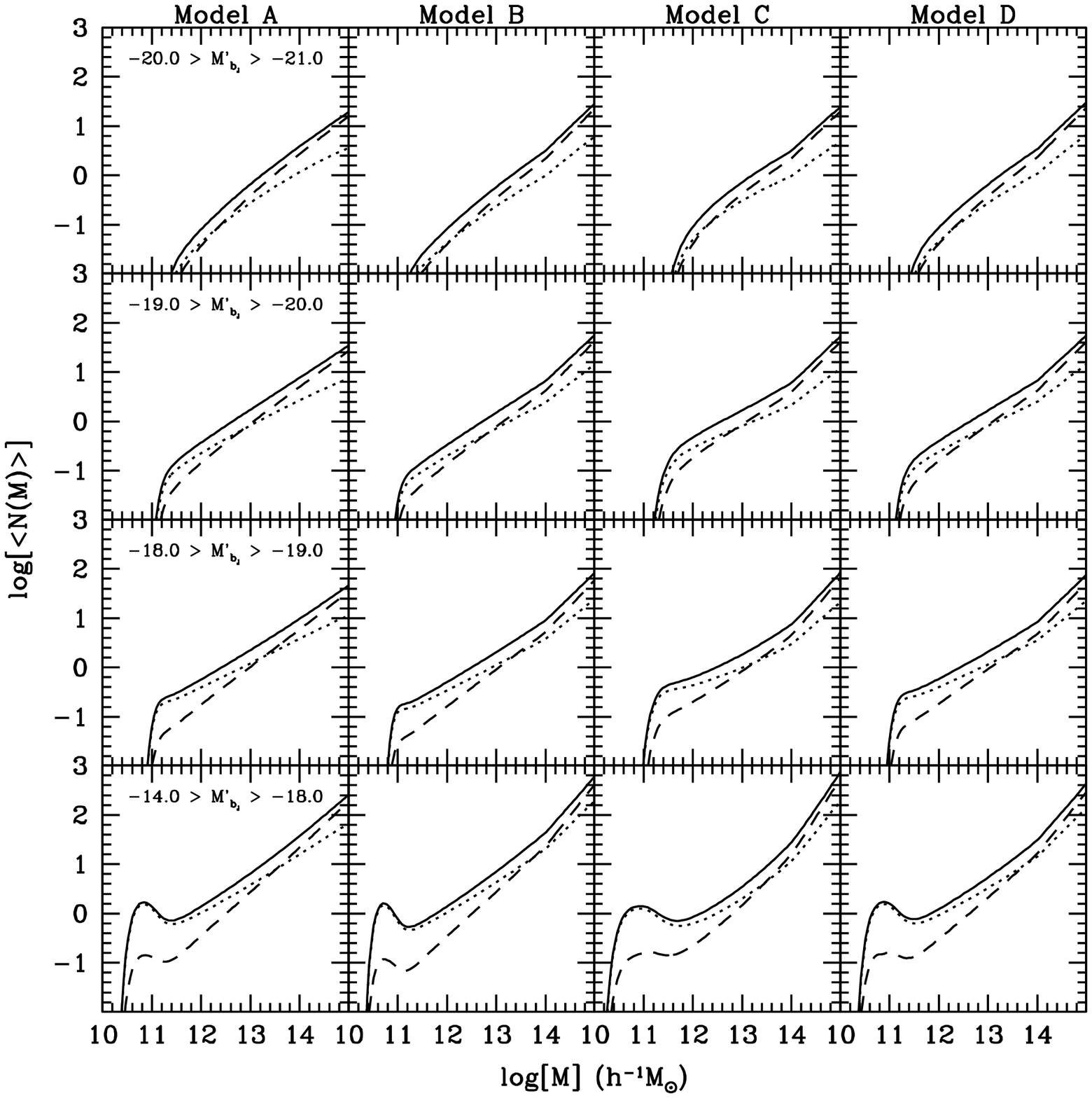,width=0.95\hdsize}}
\caption{Halo  occupation  numbers for  the  same  magnitude bins  and
models  as in  Figures~\ref{fig:xi}  and~\ref{fig:xiproj}. Dotted  and
dashed  lines  correspond  to   the  late-  and  early-type  galaxies,
respectively. See text for a detailed discussion.}
\label{fig:ngal}
\end{figure*}

In order  to  compare   our models to   the 2dFGRS   data  in a   more
quantitative  way, we proceed as follows.  For each  spectral type and
each  luminosity bin Norberg \etal (2002)  have fit a single power-law
relation of the form  $w_p(r_p)/r_p = A(\gamma) (r_p/r_0)^{\gamma}$ to
the projected correlation functions obtained from the 2dFGRS data.  We
rewrite this as
\begin{equation}
\label{linerel}
{\rm log}[w_p(r_p)/r_p] = w_0 + \gamma {\rm log}(r_p)
\end{equation}
and compute $w_0 = {\rm log}  A(\gamma) - \gamma \, {\rm log} r_0$ and
its error  from the  values of $r_0$  and $\gamma$ (and  their errors)
quoted   by  Norberg  \etal   (2002).   The   results  are   shown  in
Figure~\ref{fig:gamma} (open circles  with errorbars).  Next, for each
of  our  models  we  fit  equation~(\ref{linerel})  to  the  projected
correlation functions over the same radial interval over which Norberg
\etal (2002) fitted the data ($-0.5 \leq {\rm log}(r_p) \leq 1.3$).

Our best-fit  $w_0$ and  $\gamma$ are shown  in Figure~\ref{fig:gamma}
for all  models listed in Table~1.   Overall, the models  fit the data
reasonably well.  Although the models predict somewhat more pronounced
dependencies  of  $\gamma$  and  $w_0$  on  luminosity,  and  somewhat
shallower  correlation  functions  for  the  late-type  galaxies,  the
errorbars  on the  data are  too large  to rule  against any  of these
models.  The level of agreement  is actually quiet surprising in light
of the  oversimplified assumptions regarding  $\langle N_{\rm pair}(M)
\rangle$ and $f(r)$.  In  particular, the assumption of sub-Poissonian
occupation  distributions may  be a  poor assumption  on the  scale of
clusters (Benson \etal 2000;  Scoccimarro \etal 2001, Cooray 2002). In
addition, we  have assumed that  $f(r)$ is independent of  galaxy type
and or  luminosity (except for  the brightest galaxy, which  is always
assumed  to reside  at the  center). However,  it is  well  known that
clusters of  galaxies reveal a radial  morphology segregation (Abraham
\etal 1996; van Dokkum \etal 1998; Balogh \etal 1999), with early-type
(late-type) galaxies more confined towards the center (outer regions).
Although morphological segregation within individual haloes is easy to
take  into   account  (e.g.,   Scranton  2002a,b),  such   a  detailed
investigation is beyond the scope of this paper. However, we intend to
return to this issue in a future paper.

Figure~\ref{fig:gamma} also  shows predictions for  $\gamma$ and $w_0$
for galaxies  down to $M_{b_J} - 5  {\rm log} h =  -15.5$.  This shows
that the main  difference between the various models  occurs for faint
early-type  galaxies.  Furthermore,  whereas the  models  predict very
little luminosity dependence in  the projected correlation function of
late-type galaxies (except for  a modest increase of the normalization
with  luminosity),  for  the  early-type galaxies  a  very  pronounced
luminosity dependence is predicted.  In particular, the models predict
that early-type galaxies with $M_{b_J} - 5 {\rm log} h = -19.5$ (i.e.,
roughly  $L^{*}$), are less  strongly clustered  and with  a shallower
power-law  slope than  both brighter  and fainter  early-type galaxies
(except for model~E). Future data from the completed 2dFGRS and/or the
SDSS will  prove very useful  in further testing and  constraining the
CLF models presented here.

In summary, once assumptions are  made regarding the second moments of
the halo  occupation numbers and the spatial  distribution of galaxies
within  individual  dark  matter  haloes,  the  two-point  correlation
functions of galaxies {\it in  any magnitude interval} can be computed
from the  CLFs.  Because the fraction of  late-type galaxies decreases
with  increasing  halo  mass,  late-type galaxies  are  less  strongly
clustered  than their  early-type counterparts  at  small separations.
This  results  in  steeper  projected correlation  functions  for  the
early-type  galaxies, in  qualitative  agreement with  the data.   The
remaining discrepancies, although small,  are most likely a reflection
of  the  fact  that we  have  not  taken  the spatial  segregation  of
morphological types  in clusters into account. Finally  we stress that
accurate  correlation  functions  of  faint  early-type  galaxies  are
required  in order  to break  the  degeneracy present  in the  current
models for the CLF.

\subsection{Cross correlation functions}
\label{sec:cross}

In addition  to the (auto) correlation functions  presented above, the
CLFs  also   allow  the  computation  of   various  cross  correlation
functions. As an  example, Figure~\ref{fig:cross} plots the zero-point
$w_0$ and power-law slope $\gamma$ of the projected cross-correlations
with  bright ($-21  >  M_{b_j} -  5  {\rm log}  h  > -22$)  early-type
galaxies.      Compared    to    the     auto-correlation    functions
(Figure~\ref{fig:gamma}) the  cross correlation functions  are steeper
and  with a  higher zero-point  $w_0$. This  is due  to the  fact that
bright early-type  galaxies reside predominantly  in massive clusters.
In addition, compared to the auto-correlation functions, the late-type
galaxies  reveal   a  more   pronounced  dependence  of   $\gamma$  on
luminosity. This  is due to  the fact that clusters  are predominantly
occupied  by  early-type  galaxies:  Faint late-type  galaxies  reside
mainly  in  low-mass  haloes,  and  are  therefore  relatively  weakly
correlated  with bright early-type  galaxies. The  brightest late-type
galaxies, on the other hand,  are still predominantly found in massive
clusters,   and  therefore   strongly  correlated   with   the  bright
early-types. Note  that these predictions are likely  to depend fairly
strongly  on  the  presence  of  a radial  morphology  segregation  in
clusters,  which   is  not   taken  into  account   here.   Therefore,
Figure~\ref{fig:cross} should not be taken as a strong prediction, but
merely serves to illustrate the  power of the CLF method for computing
various statistical properties of the galaxy distribution.

\section{Halo Occupation Numbers}
\label{sec:hons}

We  now  use  the  CLFs  to  compute  halo  occupation  numbers  using
equation~(\ref{nlm}).  Results are  shown in Figure~\ref{fig:ngal} for
the   same   samples    of   galaxies   as   in   Figures~\ref{fig:xi}
and~\ref{fig:xiproj}.   Although there  are small  differences amongst
the  four models  presented here,  several  trends are  shared by  all
models.  First of all, one finds that the derivative ${\rm d}\langle N
\rangle/{\rm  d}M$ is  larger when  brighter galaxies  are considered.
Secondly,  when  sufficiently   faint  galaxies  are  included,  ${\rm
d}\langle  N \rangle/{\rm d}M$  depends more  strongly on  mass (i.e.,
$\langle N(M) \rangle$  deviates more from a single  power-law) in the
sense that  it decreases  towards lower $M$.   We can  understand this
behavior by focusing on the  contributions of the late- and early-type
galaxies.  In  all cases we  find that ${\rm d}\langle  N \rangle/{\rm
d}M$ is  largest for the early-type  galaxies.  At large  $M$ and $L$,
the majority  of galaxies are early-types, and  $\langle N(M) \rangle$
is  well fit  by a  single power-law.   At lower  $L$ the  fraction of
late-type galaxies increases, predominantly at lower masses, such that
a transition occurs from being dominated by late-type galaxies (at low
$M$)  to  being  dominated  by  early-type  galaxies  (at  high  $M$).
Finally, note that  $\langle N(M) \rangle$ is always  truncated at low
$M$, indicating the presence of a  minimum halo mass that can harbor a
galaxy of a given luminosity.  In terms of our models, this truncation
is a reflection of $\wLstar(M)$.

\subsection{Comparison with previous work}
\label{sec:compare}

Several  studies  in  the past  have  used  a  variety of  methods  to
constrain  $\langle  N(M) \rangle$  from  observations.   Jing, Mo  \&
B\"orner (1998), using the two-point correlation function and pairwise
velocity  dispersions from  the  Las Campanas  Redshift Survey  (LCRS;
Shectman  \etal  1996), obtained  $N  \propto M^{0.92}$.   Scoccimarro
\etal (2001)  fitted the measurements of counts-in-cells  from the APM
survey  (Maddox  \etal  1990)  deprojected into  three  dimensions  by
Gazta\~{n}aga (1994),  and found that $N \propto  M^{0.8}$ is required
in the high mass limit.
\begin{figure}
\centerline{\psfig{figure=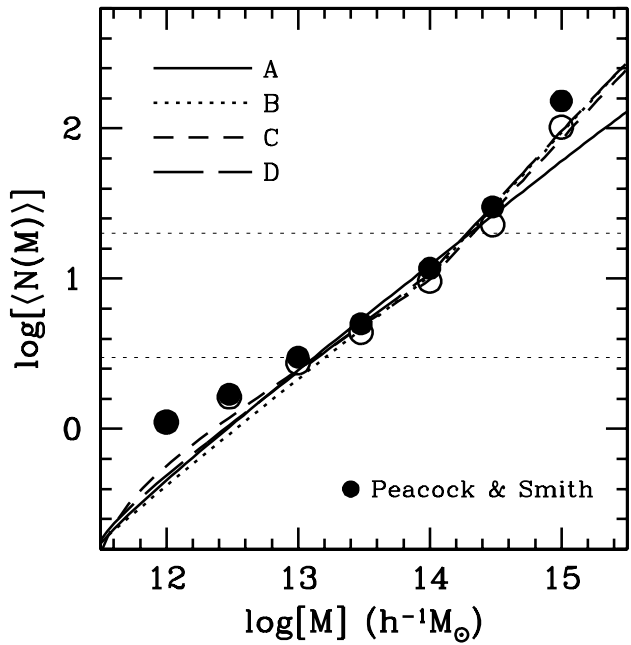,width=\hssize}}
\caption{A comparison  of   the halo  occupation  numbers  obtained by
Peacock \& Smith (2000, solid  dots) and our CLFs for  models A, B, C,
and  D  (various lines).  The    open circles correspond  to  the halo
occupation  numbers obtained after  replacing  $N$ with $N^{0.92}$  as
suggested by Peacock \& Smith for haloes at the high mass end. The two
dotted, horizontal lines indicate the range  in $N$ over which Peacock
\&  Smith were  able  to  constrain their  models {\it  directly} from
data.}
\label{fig:honps}
\end{figure}

Unfortunately, a direct comparison of these results with each other or
with  predictions from our  CLFs is  complicated for  several reasons.
First  of  all,  the  results   obtained  by  Jing  \etal  (1998)  and
Scoccimarro \etal (2001) are based on {\it apparent magnitude} limited
samples, which are difficult to  compare to halo occupation numbers in
absolute  magnitude ranges,  such  as those  obtained  from the  CLFs.
Secondly, the LCRS is based on $R$-band data for which halo-occupation
numbers are not expected to be the same as in the $b_J$-band used here
and  by  Scoccimarro et  al.  In  fact, as  shown  by  Jing \etal  the
two-point correlation  function of  the LCRS is  significantly steeper
than that of the APM survey on which the 2dFGRS data is based.

Peacock \& Smith (2000; hereafter  PS00), on the other hand, presented
halo occupation numbers for  galaxies with $M_{b_J} \leq -19.0$, which
we can compare  directly to predictions from our  CLFs.  Using data on
galaxy groups from the CfA survey (Ramella, Pisani \& Geller 1997) and
the ESO Slice Project (Ramella \etal 1999), PS00 found that the number
density  of  groups containing  $N$  galaxies  obeys $\rho(N)  \propto
N^{-2.7}$.  Equating  the number  density of groups  with multiplicity
larger than $N$ to the number  density of dark matter haloes with mass
larger than $M$, results in the halo occupation numbers shown by solid
dots  in Figure~\ref{fig:honps}.   For  comparison, we  also show  the
predictions for our  CLF models~A, B, C, and  D (various lines).  Over
the intermediate mass range $2 \times 10^{13} h^{-1} \Msun \lta M \lta
5 \times 10^{14} h^{-1} \Msun$  the PS00 estimates agree well with the
occupation numbers  obtained here. However,  at both lower  and higher
masses  the $\langle N(M)  \rangle$ of  PS00 are  significantly higher
than in any of our four  models.  The data used by PS00 was restricted
to groups  with $3  \leq N \leq  20$ (indicated by  dotted, horizontal
lines). For $N  > 20$ PS00 simply extrapolated  the power-law behavior
of $\rho(N)$ to higher multiplicity.  As PS00 argued themselves, these
results may  not be completely reliable, and  PS00 therefore suggested
to  replace  $N$  with $N^{0.92}$  at  the  high  mass end.   This  is
indicated by the open circles in Figure~\ref{fig:honps} and brings the
PS00 results in good agreement  with the $\langle N(M) \rangle$ of our
models~B, C,  and D for  haloes with $M  \gta 2 \times  10^{13} h^{-1}
\Msun$.   For $N  < 3$  the disagreement  remains. However,  here PS00
extrapolated  the power-law behavior  of $\rho(N)$  down to  $N=2$ and
obtained  the result  for  $N=1$  by matching  the  number density  of
galaxies to the observed  value.  Given the uncertainties involved, it
is encouraging  that these two results, based  on completely different
methods, agree as well as they do.

\section{Comparison with Semi-Analytical Models}
\label{sec:sam}

In  this paper we  have derived  conditional luminosity  functions for
late-  and early-type  galaxies.   We have  shown  that although  some
degeneracy remains, the overall  behavior of $\langle M/L \rangle$ and
$\langle  N(M) \rangle$  is well  constrained. We  thus  have obtained
important constraints on how galaxies have to be distributed over dark
matter haloes  in order to yield luminosity  functions and correlation
lengths  in agreement  with  the data.   The  important question  that
arises is  whether such halo  occupancy distributions can  be obtained
within the standard  framework for galaxy formation; i.e.,  how do the
cooling, feedback and star formation efficiencies depend on halo mass,
such that one can reproduce the CLFs obtained here.
\begin{figure*}
\centerline{\psfig{figure=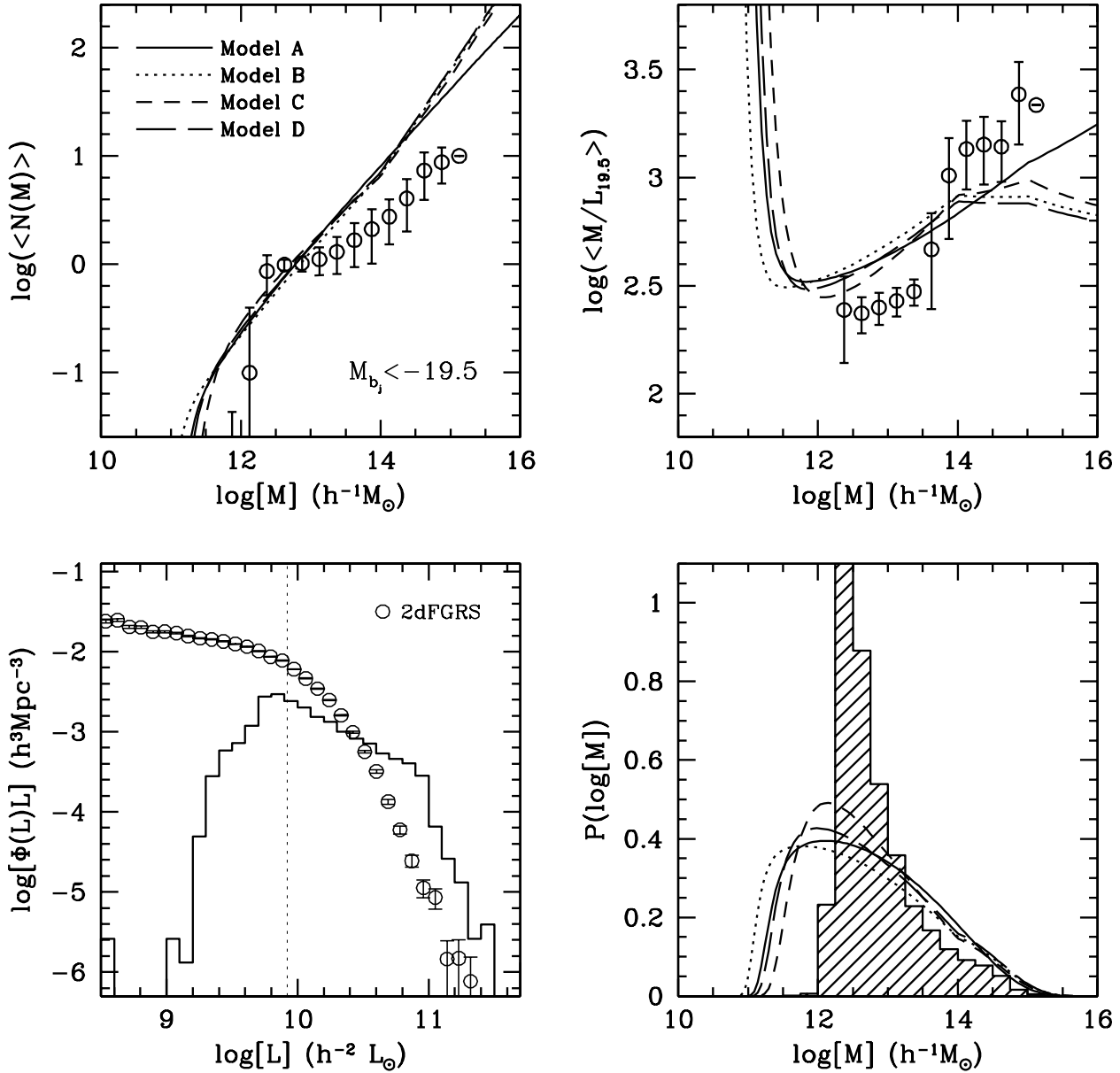,width=0.95\hdsize}}
\caption{Comparison of  the semi-analytical models  of Kauffmann \etal
(1999; K99)  with models~A, B, C,  and~D.  Upper left  panel plots the
halo occupation  numbers for galaxies with  $M_{b_J} - 5  {\rm log}h <
-19.5$.  Symbols  correspond to the semi-analytical  models, and lines
to the CLF models, as indicated.  Compared to our models, the SA-model
underpredicts the  average number of  galaxies in massive  haloes. The
upper right  panel plots the corresponding  ratio of halo  mass to the
total  luminosity  $L_{19.5}$ contributed  by  galaxies brighter  than
$M_{b_J}  - 5  {\rm log}h  = -19.5$.   The K99  model  predicts higher
(lower) values of $M/L_{19.5}$ compared  to the CLFs presented here at
the  high  (low) mass  end.   The lower  right  panel  plots the  mass
distribution of haloes hosting galaxies with $M_{b_J} - 5 {\rm log}h <
-19.5$. Again, the agreement between  the K99 model and our CLF models
is  poor,   with  the  latter   predicting  much  broader   halo  mass
distributions.   The reason  for these  various discrepancies  is that
unlike for the  CLF models presented here, the LF of  the K99 model is
strongly inconsistent with  the observed LF. This is  indicated in the
lower left panel where symbols correspond to the 2dFGRS LF of Madgwick
\etal (2002)  while the histogram corresponds  to the LF  of the model
galaxies of K99.  The vertical dotted line indicates $M_{b_J} - 5 {\rm
log}h = -19.5$.}
\label{fig:samK}
\end{figure*}

In order to address this  question we compare our results with several
semi-analytical  (SA) models in  the literature.  These SA  models use
various phenomenological prescriptions to describe the star formation,
feedback,  and  cooling  efficiencies   which  are  linked  to  merger
histories  of dark  matter haloes  obtained from  either  the extended
Press-Schechter formalism (e.g., Lacey \& Cole 1993), or directly from
numerical  simulations. Combined  with stellar  population  models and
recipes for the merging of  galaxies within merging dark matter haloes
these models yield, amongst others, luminosities of galaxies in haloes
of different masses.
\begin{figure*}
\centerline{\psfig{figure=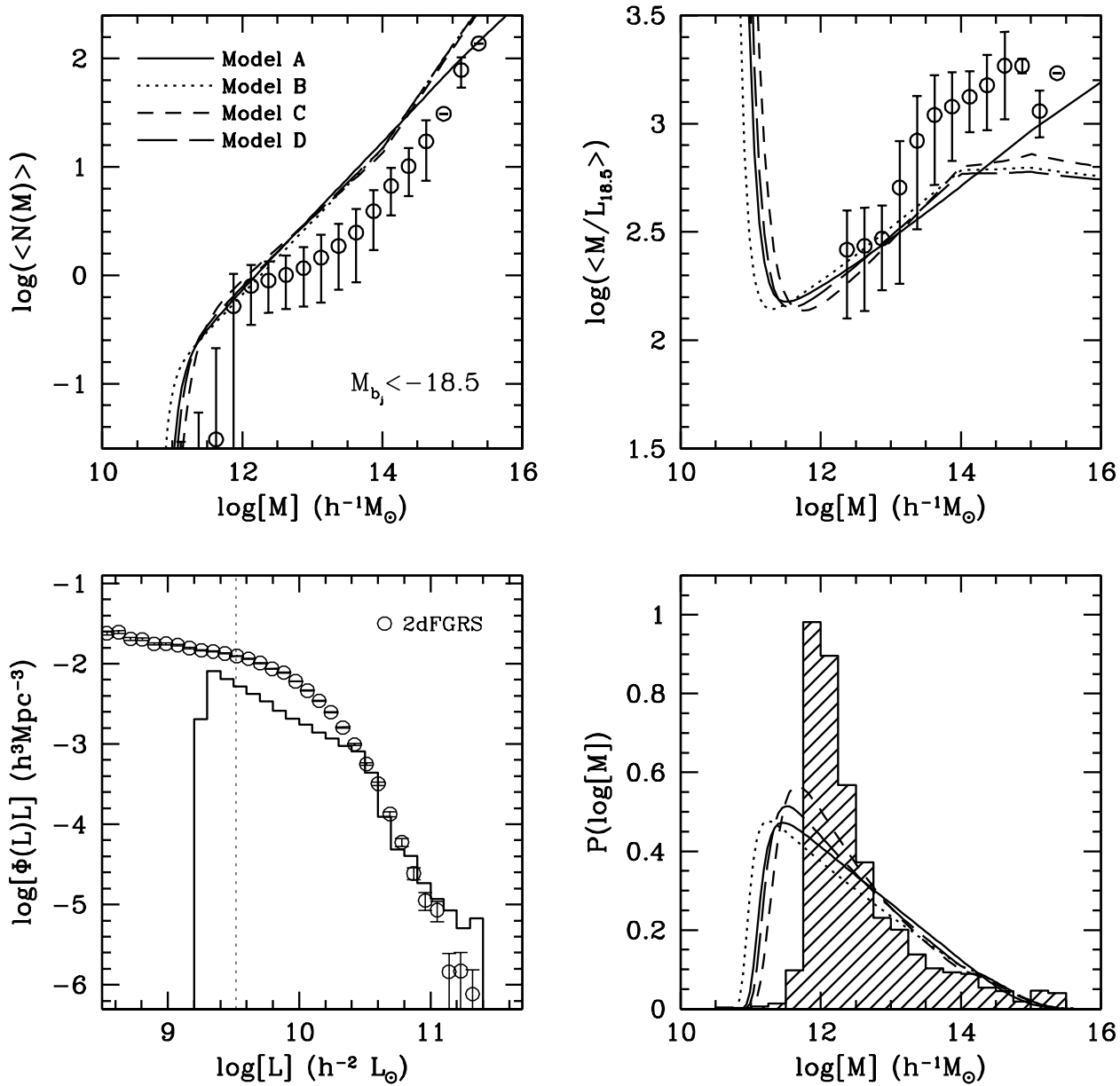,width=0.95\hdsize}}
\caption{Same  as Figure~\ref{fig:samK}, except  that here  we compare
our models  to the SA model  of Mathis \etal (2002;  M02) for galaxies
with $M_{b_J} - 5  {\rm log}h < -18.5$. Since the LF  of this SA model
is in somewhat better agreement  with the observations (see lower left
panel), the overall  agreement with our CLF models  has also improved,
though only marginally.}
\label{fig:samM}
\end{figure*}

Here  we compare  our models  with the  SA models  of  Kauffmann \etal
(1999; hereafter K99), Mathis  \etal (2002; hereafter M02), and Benson
\etal  (2002; B02)\footnote{The  models of  K99 and  M02  are publicly
available                                                            at
http://www.MPA-Garching.MPG.DE/$\sim$virgo/virgo/index-galaxy.html
while the model  of B02 was kindly provided to  us by Andrew Benson.}.
Each  of  these  three  models  used  exactly  the  same  cosmological
parameters as adopted here. K99  and M02 used numerical simulations to
compute  the merger  histories of  the dark  matter haloes,  and their
models   are  consequently   limited   in  halo   mass  by   numerical
resolution. The  model of B02, on  the other hand, is  based on merger
histories obtained  using the extended  Press-Schechter formalism (for
details see Cole  \etal 2000), and therefore allows  a comparison with
our models over  a much larger range of  halo masses and luminosities.
The  results of  K99 and  M02 are  based on  the  same semi-analytical
model,     although    the     parameters     used    are     somewhat
different. Luminosities in  the K99 model have not  been corrected for
dust extinction,  whereas those in  M02 have (though we  have verified
that  this has  only  a very  small  effect on  the results  presented
below).   Both  K99  and  M02  tuned their  model  parameters  to  fit
primarily  the zero-point  of  the Tully-Fisher  (TF) relation.   This
results in rather poor fits to  the observed LF (see below). The model
of B02 is significantly different.  It is based on the semi-analytical
models of Cole \etal (2000)  but with (i) an improved prescription for
the  effects of  tidal stripping  and dynamical  friction, and  (ii) a
model for  feedback from the photoionizing background.   Unlike in K99
and  M02, B02  tune their  parameters to  reproduce the  2dFGRS  LF of
Madgwick \etal (2002), which results in a TF zero-point that is offset
from the observed value (see discussion below).
\begin{figure*}
\centerline{\psfig{figure=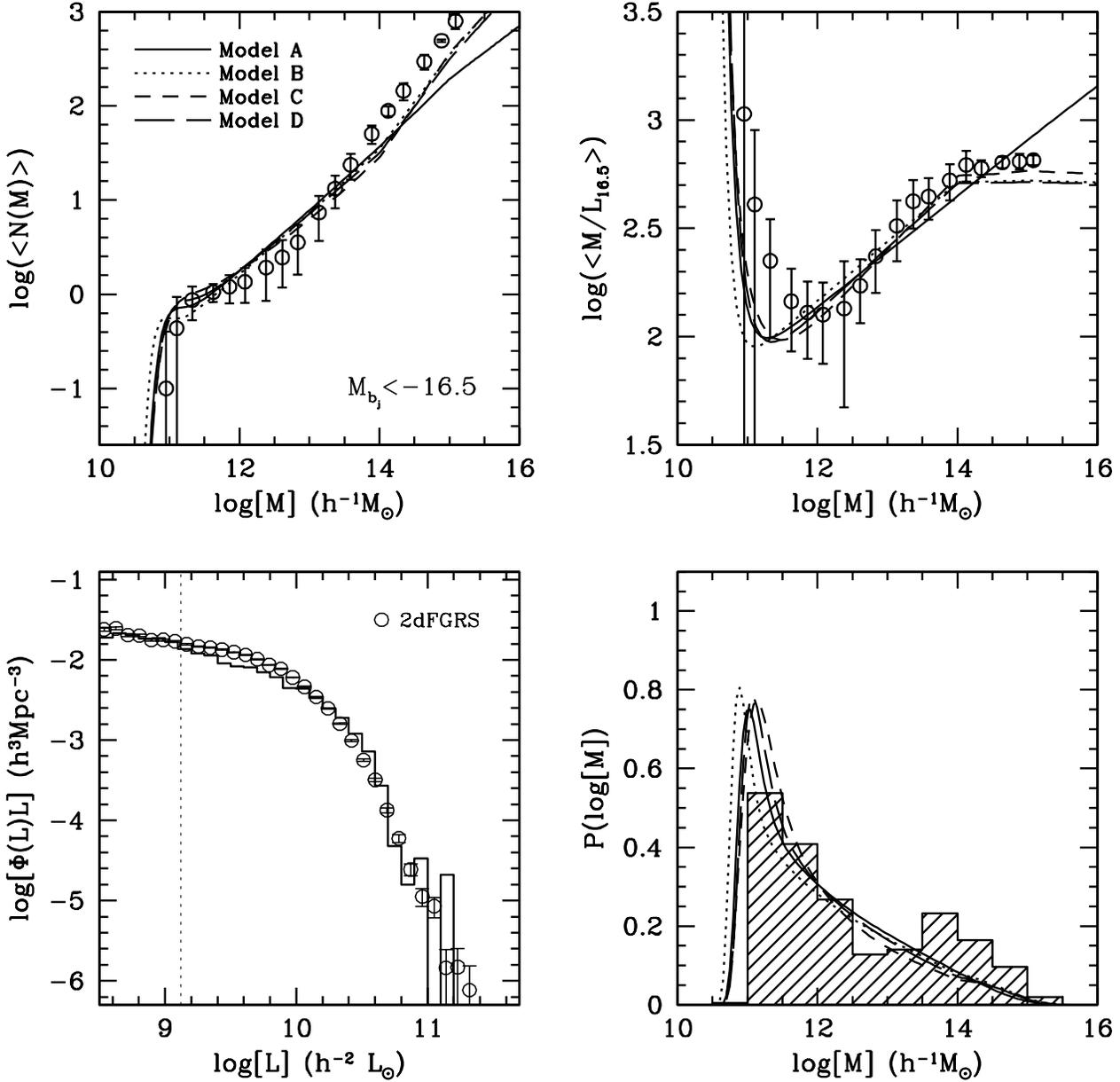,width=0.95\hdsize}}
\caption{Same as in  Figures~\ref{fig:samK} and~\ref{fig:samM} but for
the  semi-analytical model of  Benson \etal  (2002; B02)  for galaxies
with $M_{b_J}  - 5 {\rm log}h <  -16.5$. Contrary to those  of K99 and
M02, this  model agrees  well with  both our CLF  models and  with the
observed  2dFGRS  LF.   Since  B02  tuned their  model  parameters  to
reproduce  this  particular  LF,  the   latter  does  not  come  as  a
surprise. However, the good agreement for the halo occupation numbers,
mass-to-light ratios, and halo  mass distributions with our CLF models
is  remarkable   given  the   completely  different  nature   of  both
techniques.  The  remaining disparities between  the SA model  and our
CLF models  can be contributed  to the remaining  disagreement between
the observed LF and that of the B02 model.}
\label{fig:samB}
\end{figure*}

Before we  can make a  fair comparison we  need to adopt  a consistent
definition of halo  masses. Both the SA models of  K99 and M02 provide
two measures  for the halo  mass associated with each  galaxy; $M_{\rm
FOF}$,   the  mass   of   the  halo   obtained   using  the   standard
friends-of-friends (FOF)  algorithm (Davis \etal 1985)  with a linking
length  of  $b=0.2$  times  the mean  inter-particle  separation,  and
$M_{200}$, the mass of a sphere centered on the most-bound-particle of
the FOF group with a radius inside of which the average density is 200
times the  {\it critical} density.   For each individual  halo $M_{\rm
FOF}$  and  $M_{200}$  can  differ substantially,  and  therefore  the
results are sensitive  to which halo mass we  adopt.  Since our models
rely  on the  Sheth \&  Tormen (1999)  halo mass  function,  and since
Jenkins \etal  (2001) have shown that  this is in  very good agreement
with the mass function of FOF groups with a linking length of $b=0.2$,
we  use $M_{\rm  FOF}$.  B02  define their  halos as  spheres  with an
average  overdensity of  320. We  convert these  masses  to $M_{180}$,
consistent  with the  definition  of halo  mass  used throughout  this
paper,  assuming  that  dark  matter  halos  follow  the  NFW  density
distribution  (Navarro,  Frenk \&  White  1992)  with a  concentration
parameter given by Eke, Navarro  \& Steinmetz (2001).  Next we need to
be careful about the use of luminosities.  The SA models give for each
galaxy the absolute  magnitudes in the $B$ and $V$  bands. In order to
allow a  fair comparison with our  models, we compute  for each galaxy
$M_{b_J}  - 5  {\rm log}h$  using $b_J  = B  - 0.28  (B-V)$  (Blair \&
Gilmore  1982) and  adopting the  Hubble constant  used in  the models
($h=0.7$).

We first  compare our results  with the model  of K99. Because  of the
numerical resolution of the simulation used by K99, only galaxies with
a  stellar  mass  $M_{*}  \geq  1.4 \times  10^{10}  h^{-1}\Msun$  are
available.  Although this includes galaxies  down to $M_{b_J} - 5 {\rm
log}h \simeq -18$, the sample  of model galaxies is only complete down
to $M_{b_J} - 5 {\rm log}h = -19.5$.  When comparing for instance halo
occupation numbers,  completeness is important and  we therefore limit
ourselves  here to a  comparison of  galaxies with  $M_{b_J} -  5 {\rm
log}h  \leq -19.5$.   The  upper left  panel of  Figure~\ref{fig:samK}
plots the number of galaxies with  $M_{b_J} - 5 {\rm log}h \leq -19.5$
as function of  halo mass.  Open circles correspond  to the K99 model,
with errorbars indicating  the rms scatter.  In addition,  we show the
predictions for  models A,  B, C, and  D (various  lines). Differences
between these four models are typically small. For $M > 10^{13} h^{-1}
\Msun$  the SA model  predicts fewer  galaxies per  halo than  our CLF
models, while  the opposite is true  for haloes with $M  \sim 5 \times
10^{12} h^{-1}  \Msun$.  The upper  panel on the right  plots $\langle
M/L_{19.5}  \rangle$ as  function of  halo mass.   Here  $L_{19.5}$ is
defined as  the total  luminosity of galaxies  with $M_{b_J} -  5 {\rm
log}h \leq  -19.5$ in each individual  halo.  For $M  > 10^{14} h^{-1}
\Msun$ the SA model overpredicts $\langle M/L_{19.5} \rangle$ compared
to our CLF models, in agreement  with the fact that K99 predicts lower
occupation  numbers.  However,  at  $M \simeq  10^{13} h^{-1}  \Msun$,
where both  the SA and  CLF models predict  on average one  galaxy per
halo,  the  former predicts  significantly  lower $\langle  M/L_{19.5}
\rangle$.  This  indicates that the average luminosity  of galaxies in
haloes with $M=10^{13} h^{-1} \Msun$ is higher in the SA model than in
our  CLF  models.  Finally,  the  lower  right  panel plots  the  mass
distribution of haloes harboring galaxies with $M_{b_J} - 5 {\rm log}h
\leq -19.5$. Note that a  halo containing $N$ such galaxies is counted
$N$ times.  The hatched histogram corresponds to the SA model, whereas
the various lines show predictions for our CLF models computed from
\begin{equation}
\label{massprob}
P(M) {\rm d}M = {\int_{L_{\rm min}}^{\infty} \Phi(L \vert M) {\rm d}L 
\over \int_{L_{\rm min}}^{\infty} \Phi(L) {\rm d}L} \, n(M) \, {\rm d}M.
\end{equation}
with $L_{\rm min}$ the lower luminosity limit of the sample. Again the
agreement between  the SA and CLF  models is extremely  poor, with the
latter predicting  much broader mass distribution  functions.  None of
these disagreements, however, come as  a surprise. After all, as shown
in the lower left  panel, the K99 model predicts a LF  that is in very
poor agreement with the observed LF  of the 2dFGRS, and thus also with
the LFs corresponding to our CLF models.

Figure~\ref{fig:samM} shows  a similar comparison but now  with the SA
model  of  M02. Since  the  numerical  simulations  used by  M02  have
somewhat  higher numerical  resolution, we  can extend  our comparison
down  to  $M_{b_J} -  5  {\rm log}h  =  -18.5$.   Mathis \etal  (2002)
focussed  on  modeling  the   {\it  local}  galaxy  population,  using
numerical  simulations  constrained to  reproduce  the local  smoothed
linear density  field out  to a distance  of $8000 \kms$.   Since this
corresponds to a cosmological volume that is smaller than that in K99,
the errorbars are  somewhat larger. As shown in  the lower left panel,
the M02  model significantly improves the agreement  with the observed
LF at  the bright-end compared to  K99, even though  it still severely
underpredicts the  number of  galaxies with $M_{b_J}  - 5 {\rm  log} h
\gta  -20$.  This explains  why  the  SA  model predicts  (i)  smaller
occupation  numbers, (ii)  higher  mass-to-light ratios,  and (iii)  a
higher mean halo  mass for galaxies with $M_{b_J} -  5 {\rm log}h \leq
-18.5$ compared to our CLF models.

Finally,  in Figure~\ref{fig:samB}  we compare  our models  to  the SA
model of B02.   Since numerical resolution does not  play a role here,
we extend the comparison down to galaxies with $M_{b_J} - 5 {\rm log}h
= -16.5$.   As shown in the lower  left panel, the B02  model yields a
much  better  fit to  the  observed  LF than  either  the  K99 or  M02
models. Since  B02 chose  to tune their  model parameters to  fit this
particular LF,  rather than  the zero-point of  the TF  relation, this
does not come  as a surprise. The good agreement  with our CLF models,
however, is not a trivial result. Although some amount of disagreement
remains, this  can all be  attributed to the  fact that the  B02 model
slightly  over(under)-predicts  the  LF  at  luminosities  just  above
(below) $L^{*}$.   All in all, the agreement  is extremely encouraging
and seems  to suggest that our  CLFs correspond to  a galaxy formation
scenario that is very similar to the SA model of B02.

In  summary,  our  CLF   models  imply  halo  occupation  numbers  and
mass-to-light   ratios   that  are   in   poor   agreement  with   the
semi-analytical  models  of K99  and  M02,  but  in surprisingly  good
agreement with the model of B02.  However, this does not mean that the
B02 models are better or more reliable than the models of K99 and M02.
Rather, the differences  between the various SA models  reflect a well
known  problem in  galaxy formation  theory, namely  the  inability of
models to simultaneously fit the  observed LF and zero-point of the TF
relation  (e.g.,   Kauffmann  \etal  1993;  Cole   \etal  1994,  2000;
Somerville \& Primack 1999; Benson \etal 2000). In fact, this does not
seem  a  problem related  to  the  particular  prescriptions for  star
formation and/or feedback used in  these models, but to reflect a more
fundamental problem  related to the particular  cosmological model: as
we have  shown in detail  in Paper~1, even  the CLF models,  which are
independent  of  how  galaxies  form,  are inconsistent  with  the  TF
zero-point. Since  both the  SA model  of B02 and  our CLF  models are
tuned to  fit the LF, one  should not expect  good agreement regarding
halo occupation numbers  with the K99 and M02  models, which are tuned
to fit the  TF zero-point.  However, it is  extremely encouraging that
the SA  model that  {\it does} fit  the observed LF  agrees remarkably
well with  our CLF models.  The  fact that these  two wildly different
techniques imply  consistent mass-to-light ratios  and halo occupation
numbers,  is an important  step forward  in our  quest for  a coherent
picture  of galaxy  formation within  a CDM  cosmology.   The question
whether  the occupation distributions  inferred from  our CLFs  can be
made consistent with the  standard framework for galaxy formation thus
seems to have a positive answer.

\section{Conclusions}
\label{sec:concl}

Data from large  ongoing redshift surveys such as  the SDSS and 2dFGRS
provide  a  wealth of  information  on  the  clustering properties  of
galaxies  as  function  of  morphological  type,  luminosity,  surface
brightness,  star formation rate,  etc.  At  the same  time, numerical
simulations and  sophisticated analytical techniques  are continuously
improving  our  understanding of  the  clustering  properties of  dark
matter  haloes within CDM  cosmologies. The  ultimate challenge  is to
constrain  how  galaxies  with different  properties occupy  haloes of
different  masses.   This  information  not  only  gives  an  easy  to
interpret description  of the galaxy-dark matter  connection, but also
yields important constraints on galaxy formation theories.

In  this paper we  have used  recent observations  from the  2dFGRS to
constrain the  CLFs of early-  and late-type galaxies.   Although some
amount  of degeneracy remains,  the CLFs  are well  constrained.  They
indicate that  the average mass-to-light ratios of  dark matter haloes
have a minimum  of $\sim 100 h  \MLsun$ around a halo mass  of $\sim 3
\times  10^{11}  h^{-1} \Msun$.   Towards  lower  masses $\langle  M/L
\rangle$ increases  rapidly, and matching  the faint-end slope  of the
observed LF requires  that haloes with $M <  10^{10} h^{-1} \Msun$ are
virtually devoid of  galaxies producing light.  At the  high mass end,
the observed  clustering properties of galaxies  require that clusters
have mass-to-light ratios (in the photometric $b_J$ band) in the range
of roughly $500 h \MLsun$ to  $1000 h \MLsun$.  Finally, the fact that
early-type  galaxies  are   more  strongly  clustered  than  late-type
galaxies  requires  that  the  fraction  of late-type  galaxies  is  a
strongly declining function of  halo mass, although the exact relation
is currently  poorly constrained. Forthcoming data from  both the SDSS
and the 2dFGRS will reduce  the errors on the current measurements and
provide correlation lengths down  to fainter magnitudes, both of which
are essential to further constrain the CLFs.

The halo  occupation numbers $\langle  N(M) \rangle$ implied  by these
CLFs  depend rather  strongly on  the magnitude  interval  of galaxies
considered.  Typically,  including fainter galaxies  results in weaker
mass  dependencies and a  stronger departure  from a  single power-law
form.  This  is mainly  a reflection  of the way  in which  early- and
late-type  galaxies contribute: early-type  galaxies reveal  a steeper
$\langle N(M) \rangle$ than  the late-type galaxies.  When only bright
galaxies  are considered,  $\langle N(M)  \rangle$  is well  fit by  a
single  power-law  (with a  low  mass  cut-off)  and is  dominated  by
early-type  galaxies. When sufficiently  faint galaxies  are included,
the late-type  galaxies start to dominate  at the low  mass end. Since
these  reveal  a  shallower  $\langle N(M)  \rangle$  than  early-type
galaxies,   the  shape   of  $\langle   N(M)  \rangle$   becomes  more
complicated.  We have compared  our halo occupation numbers with those
obtained by  Peacock \& Smith  (2000) from the multiplicity  of galaxy
groups. Except for  haloes with $\langle N \rangle  \lta 3$, where the
results  of   Peacock  \&  Smith  are  uncertain   due  to  unreliable
extrapolation, the agreement is remarkably good.

Once  assumptions  are  made  about  the second  moment  of  the  halo
occupation  number  distributions  and  the  spatial  distribution  of
galaxies  within   individual  haloes,  the   two-point  galaxy-galaxy
correlation function can be computed  from the CLFs presented here for
any magnitude bin and for  both early- and late-type galaxies (as well
as  their cross  correlation). We  presented a  number  of correlation
functions  under   the  assumptions  that   (i)  $P(N  \vert   M)$  is
sub-Poissonian, (ii)  the brightest galaxy always resides  at the halo
center, and (iii) the  remaining satellite galaxies follow the density
distribution of the  dark matter. In order to  facilitate a comparison
with real  data we also computed the  projected correlation functions.
For late-type  galaxies, we  find that the  power-law slopes  of these
projected   correlation  functions   are   virtually  independent   of
luminosity.   For early-type  galaxies, on  the other  hand,  a fairly
strong  luminosity dependence is  predicted; early-type  galaxies with
luminosities around  $L^{*}$ are less  strongly clustered, and  with a
shallower power-law slope,  than early type galaxies with  either $L >
L^{*}$  or $L  <  L^{*}$.  A  direct  comparison with  the slopes  and
zero-points of  the projected correlation functions  obtained from the
2dFGRS reveals  good agreement. We also presented  predictions for the
cross     correlation     functions     with     bright     early-type
galaxies.  Typically,   these  are  stronger  and   steeper  than  the
corresponding   auto-correlation   functions,   which  is   a   direct
consequence of  the fact that bright  early-types reside predominantly
in massive clusters.

Finally  we have  compared halo  occupation numbers  and mass-to-light
ratios inferred  from our CLFs  with predictions from  three different
semi-analytical models  for galaxy formation.   Overall, the agreement
between our models and the SA  models of both K99 and M02 is extremely
poor.  However,  the reason for  this discrepancy is  well understood.
Both K99 and M02 have normalized their models to fit the zero-point of
the TF  relation.  It  is well known,  that current models  for galaxy
formation  fail  to  simultaneously  fit  the TF  zero-point  and  the
observed  LF of  galaxies.   In  fact, exactly  the  same problem  was
identified in Paper~1  based on modeling the CLF.   Indeed, the models
of K99 and M02 yield LFs that differ strongly from that of the 2dFGRS,
and  thus from  those of  our CLF  models.  It  should not  come  as a
surprise that therefore the  halo occupation numbers and mass-to-light
ratios do  not match.   Benson \etal (2002),  however, presented  a SA
model  that  was  normalized  to  fit the  observed  2dFGRS  LF.   The
agreement  of  this  particular  SA  model  with  our  CLF  models  is
remarkably  good.   Given  the  completely different  nature  of  both
techniques  to  predict  halo  occupation  numbers  and  mass-to-light
ratios, this  agreement is  far from trivial.   In fact,  it indicates
that the  technique used here has recovered  a statistical description
of  how galaxies  populate dark  matter haloes  which is  not  only in
perfect agreement  with the  data, but which  in addition  fits nicely
within the standard framework for galaxy formation.


\section*{Acknowledgements}

We  are grateful  to  Andrew Benson,  Guinevere  Kauffmann and  Hugues
Mathis  for  making  the  results from  their  semi-analytical  models
available,  and to  Robert Smith,  John Peacock  and Carlos  Frenk for
valuable discussion.   XY thanks the MPG-CAS  student exchange program
for  financial support,  and FvdB  thanks the  Institute  for Advanced
Study  in  Princeton  and  the  Department  of  Physics  at  New  York
University for their hospitality during visits in October 2002.



\appendix

\section[]{The two-point correlation function}
\label{sec:AppA}

The  two-point correlation  function of  the  evolved, non-linear dark
matter mass distribution is given by
\begin{equation}
\label{xidmmass}
\xi_{\rm dm}(r) = \int_{0}^{\infty} {{\rm d}k \over k} \, 
\Delta_{\rm  NL}^2(k) \, {\sin kr \over kr}
\end{equation}
Here
\begin{equation}
\label{deltaNL}
\Delta_{\rm  NL}^2(k) = {1 \over 2 \pi^2} \, k^3 \, P_{\rm NL}(k)
\end{equation}
is the  dimensionless form  of the evolved,  non-linear power-spectrum
$P_{\rm NL}(k)$.   Throughout this paper  we use the  fitting function
for $\Delta_{\rm  NL}^2(k)$ given by  Smith \etal (2002),  and compute
$\xi_{\rm dm}(r)$ using equation~(\ref{xidmmass}).

For what follows, we split this correlation function in two parts:
\begin{equation}
\label{dmcorrsplit}
\xi_{\rm dm}(r) = \xi_{\rm dm}^{1 {\rm h}}(r) +
                  \xi_{\rm dm}^{2 {\rm h}}(r) \, .
\end{equation}
Here $\xi_{\rm dm}^{1  {\rm h}}$ corresponds  to pairs within the same
halo (the ``1-halo'' term), while $\xi_{\rm dm}^{2 {\rm h}}$ describes
the correlation between  dark  matter particles that  occupy different
haloes (the ``2-halo'' term).  The 1-halo term is given by
\begin{equation}
\label{xidm1h}
\xi_{\rm dm}^{1 {\rm h}}(r) = \int_{0}^{\infty} {\rm d}k \, k^2 \, 
{\sin kr \over kr} \, \int_{0}^{\infty} {\rm d}M \, n(M) \,
\left[ \hat{\delta}(M;k) \right]^2
\end{equation}
where
\begin{equation}
\label{FTdensprof}
\hat{\delta}(M;k) =  \int_{0}^{r_v} {\rho(r) \over \bar{\rho}} \,
{\rm e}^{-i{\bf  k}\cdot{\bf r}}  {\rm d}^3{\bf r}
\end{equation}
is  the  Fourier transform  of  the  halo  density profile  $\rho(r)$,
truncated  at the  virial radius  $r_v$ (e.g.,  Neyman \&  Scott 1952;
McClelland  \& Silk  1977; Seljak  2000; Ma  \& Fry  2000; Scoccimarro
\etal 2001).  We assume that $\rho(r)$ has the NFW form
\begin{equation}
\label{NFW}
\rho(r) = \frac{\bar{\delta}\bar{\rho}}{(r/r_{\rm s})(1+r/r_{\rm
 s})^{2}},
\end{equation}
where $r_s$  is a characteristic  radius, $\bar{\rho}$ is  the average
density  of  the  Universe,  and  $\bar{\delta}$  is  a  dimensionless
amplitude which  can be expressed  in terms of the  halo concentration
parameter $c=r_v/r_s$ as
\begin{equation}
\label{overdensity}
\bar{\delta} = {180 \over 3} \, {c^{3} \over {\rm ln}(1+c) - c/(1+c)}.
\end{equation}
(cf., Navarro, Frenk \&  White 1997).  Numerical simulations show that
$c$ is  correlated with halo  mass, and we  use the relation  given by
Eke, Navarro \& Steinmetz (2001), converted to the $c$ appropriate for
our   definition   of    halo   mass.    Substitution   of~(\ref{NFW})
in~(\ref{FTdensprof}) yields, after some algebra,
\begin{equation}
\label{xidmanal}
\hat{\delta}(M;k) = 4 \pi \bar{\delta} r_s^3 \left[ 
{\cal C} + {\cal S} - {\sin(kr_v) \over kr_s + kr_v} \right]
\end{equation}
(Scoccimarro \etal 2001). Here
\begin{equation}
{\cal C} = \cos(kr_s) [{\rm Ci}(kr_s + kr_v) - {\rm Ci}(kr_s)]
\end{equation}
\begin{equation}
{\cal S} = \sin(kr_s) [{\rm Si}(kr_s + kr_v) - {\rm Si}(kr_s)]
\end{equation}
with  ${\rm Ci}(x)  = -\int_{x}^{\infty}  {\rm d}t  \,  \cos(t)/t$ and
${\rm Si}(x) = \int_{0}^{x} {\rm d}t \, \sin(t)/t$ the cosine and sine
integrals, respectively.

In order to  compute the galaxy-galaxy two-point correlation function,
$\xi_{\rm gg}(r)$ we use the same 1-halo and 2-halo split as before:
\begin{equation}
\label{ggcorrsplit}
\xi_{\rm gg}(r) = \xi_{\rm gg}^{1 {\rm h}}(r) +
                  \xi_{\rm gg}^{2 {\rm h}}(r) \, .
\end{equation}
For the 1-halo term we need to specify the distribution of galaxies in
individual haloes, while  the 2-halo term is  given by the correlation
of the population of dark matter haloes.

Consider all galaxies with luminosities in the range $[L_1, L_2]$, and
let $\xi_{\rm gg}(r)$ correspond to the two-point correlation function
of this subset  of galaxies. By definition, the  total number of pairs
of galaxies  {\it per comoving  volume} in this luminosity  range that
have separations in the range $r \pm {\rm d}r/2$ is
\begin{equation}
\label{corrdef}
n_{\rm  pair} =  {\overline{n}_{\rm g}^2  \over 2}  \, [1  + \xi_{\rm
gg}(r)] \, 4 \pi r^2 {\rm d}r.
\end{equation}
where the factor $1/2$ corrects for double counting of each pair. Here
$\overline{n}_{\rm g}$ is the mean  number density of galaxies with $L
\in [L_1, L_2]$ which is given by
\begin{equation}
\label{barng}
{\overline n}_{\rm g} = \int_{0}^{\infty} n(M) \, \langle N(M)\rangle
\, {\rm d}M \,,
\end{equation}
and $\langle  N(M) \rangle$, given by  equation~(\ref{nlm}), gives the
mean number of  galaxies in the specified luminosity  range for haloes
of mass $M$.

Consider a galaxy with $L \in [L_1, L_2]$ that lives in a halo of mass
$M_1$.  There  are, on average,  $n(M_2) \, 4  \pi r^2 {\rm  d}r$ dark
matter haloes with mass $M_2$ that are separated from this galaxy by a
distance in  the range $r  \pm {\rm d}r/2$.   In each of  these haloes
there are on average $\langle  N(M_2) \rangle$ galaxies that also have
a luminosity in the range $[L_1,L_2]$.  Integrating over the halo mass
function,  and taking  account of  the clustering  properties  of dark
matter haloes,  we obtain  the {\it excess}  number $N_{+}$  of galaxy
pairs {\it per galaxy in a halo of mass} $M_1$ with separations in the
range $r\pm {\rm d}r/2$:
\begin{eqnarray}
\label{nexc}
\lefteqn{N_{+}(M_1) = 4 \pi r^2 {\rm d}r} \nonumber \\
& & \times \int_{0}^{\infty} n(M_2) \,\langle N(M_2)\rangle \, \xi_{\rm hh}(r;
M_1, M_2) \, {\rm d}M_2.
\end{eqnarray}
Here $\xi_{\rm hh}(r; M_1, M_2)$ is the cross correlation between dark
matter haloes of mass $M_1$ and $M_2$. Note  that we implicitly assume
that  all galaxies in  a  given halo are  located  at the halo center.
Since  the  separation  between individual  haloes   is typically much
larger  than the separation  between  galaxies in  the same halo, this
simplification does not influence the results.

Multiplying $N_{+}(M_1)$ with $\langle N(M_1) \rangle$ and integrating
over the  halo mass function gives  the total excess  number of galaxy
pairs  {\it per  comoving  volume} with  separations  in the  required
range.  Combining this with equation~(\ref{corrdef}) yields
\begin{equation}
\label{fbarn}
{\overline n}_{\rm g}^2 \, \xi_{\rm gg}^{\rm 2h}(r) \, 
4 \pi r^2 {\rm d}r = \int_{0}^{\infty} n(M_1) 
\langle N(M_1) \rangle \, N_{+}(M_1) \, {\rm d}M_1.
\end{equation}

What  remains  is to specify     the dark halo  correlation   function
$\xi_{\rm hh}(r; M_1,M_2)$.   Mo  \& White (1996)  developed  a model,
based  on  the Press-Schechter  formalism, that  describes the bias of
dark matter  haloes of mass  $M$ with respect to  the dark matter mass
distribution. According to this,
\begin{equation}
\label{xihalo}
\xi_{\rm hh}(r;M_1,M_2) = b(M_1) \, b(M_2) \, \xi_{\rm dm}^{\rm 2h}(r),
\end{equation}
We use the functional form of  $b(M)$ suggested by Sheth, Mo \& Tormen
(2001),  which takes account  of ellipsoidal  collapse, and  which has
been shown to accurately match the correlation function of dark matter
haloes in $N$-body simulations (Jing 1998; Sheth \& Tormen 1999):
\begin{eqnarray}
\label{bm}
b(M) & = & 1 + {1\over\sqrt{a}\delta_{c}(z)}
\Bigl[ \sqrt{a}\,(a\nu^2) + \sqrt{a}\,b\,(a\nu^2)^{1-c} - \nonumber \\
& & {(a\nu^2)^c\over (a\nu^2)^c + b\,(1-c)(1-c/2)}\Bigr],
\end{eqnarray}
with  $a=0.707$, $b=0.5$, $c=0.6$  and $\nu  = \delta_c  / \sigma(M)$.
This bias  is defined for haloes  at $z=0$. In  Appendix~B we describe
how to compute the bias of haloes and galaxies at $z > 0$.

Substituting         equations~(\ref{xihalo})         and~(\ref{nexc})
in~(\ref{fbarn}) yields
\begin{equation}
\label{xi2h}
\xi_{\rm gg}^{\rm 2h}(r) = \overline{b}^2 \, \xi_{\rm dm}^{\rm 2 h}(r) ,
\end{equation}
with
\begin{equation}
\label{avbias}
\overline{b} = {1 \over \overline{n}_{\rm g}} \int_{0}^{\infty} n(M) \,
\langle N(M) \rangle \, b(M) \, {\rm d}M\,.
\end{equation}
Equations~(\ref{xi2h}) and~(\ref{avbias})  specify the 2-halo  term of
$\xi_{\rm gg}(r)$.  Using a similar analysis as above, one obtains the
1-halo term of the galaxy correlation function as
\begin{equation}
\label{xi1h}
{{\overline n}_{\rm g}^2 \over 2} \, \xi_{\rm gg}^{\rm 1h}(r) 
4 \pi r^2 {\rm d}r = {\rm d} r \int_{0}^{\infty} 
n(M) \langle N_{\rm pair}(M)\rangle f(r)\, {\rm d} M,
\end{equation}
where  $\langle N_{\rm  pair}(M)\rangle$ is  the mean  number  of {\it
pairs}  in haloes  of mass  $M$,  and $f(r)$  gives the  corresponding
distribution  of pair  separations (e.g.,  Berlind \&  Weinberg 2002).
Neither $\langle  N_{\rm pair}(M)\rangle$  nor $f(r)$ can  be obtained
from  the  CLF.   In  order  to compute  $\xi_{\rm  gg}^{\rm  1h}(r)$,
additional    assumptions   have   therefore    to   be    made   (see
Section~\ref{sec:corrfunc}).   The  1-halo term,  on  the other  hand,
depends  only on  $\langle  N(M)\rangle$ and  is therefore  completely
specified by the CLF.

\section[]{The effective bias of galaxies at non-zero redshift}
\label{sec:AppB}

Dark matter haloes at redshift $z$ are biased with respect to the dark
matter mass distribution at $z$ according to
\begin{equation}
\label{appBa}
\xi_{\rm hh}(r,z;M_1,M_2) = b(M_1,z) \, b(M_2,z) \, \xi_{\rm dm}^{\rm 2h}(r,z)
\end{equation}
For  this same  population  of haloes  one  can define  a second  bias
according to
\begin{equation}
\label{appBb}
\xi_{\rm  hh}(r,0;M_1,M_2)  =  b_0(M_1,z)  \, b_0(M_2,z)  \,  \xi_{\rm
dm}^{\rm 2h}(r,0)
\end{equation}
where $b_0(M,z)$ describes the bias  at $z=0$ of the {\it descendants}
of haloes  that have mass $M$ at  redshift $z$ (Fry 1996;  Mo \& White
1996, 2002).  $b_0(M,z)$ can be  expressed in terms of the bias $b(M)$
at $z=0$ given by equation~(\ref{bm}) as
\begin{equation}
\label{appBc}
b_0(M,z) = 1 + {D(0) \over D(z)} \, \left[ b(M) - 1 \right]
\end{equation}
(which is  accurate to leading  order), with $D(z)$ the  linear growth
rate at redshift $z$. On sufficiently large scales, which are still in
the linear regime,
\begin{equation}
\label{appBd}
\xi_{\rm hh}(r,z;M_1,M_2) = \left({D(z) \over D(0)}\right)^2 \, 
\xi_{\rm hh}(r,0;M_1,M_2)
\end{equation}
so that
\begin{equation}
\label{appBe}
\xi_{\rm hh}(r,z;M_1,M_2) = b_{\rm eff}(M_1,z) \, b_{\rm eff}(M_2,z) 
\, \xi_{\rm dm}^{\rm 2h}(r,0)
\end{equation}
with
\begin{equation}
\label{appBf}
b_{\rm eff}(M,z) = {D(z) \over D(0)} + b(M) - 1
\end{equation}
Substituting $b(M)$ in equation~(\ref{avbias}) with $b_{\rm eff}(M,z)$
one obtains the effective bias  of {\it galaxies} at redshift $z$ with
luminosities in the range $L_1 < L < L_2$:
\begin{equation}
\label{appBg}
\overline{b}_{\rm eff}(z) \equiv \left( \xi_{\rm gg}^{\rm 2h}(r,z) 
\over \xi_{\rm dm}^{\rm 2h}(r,0) \right)^{1/2} =
{D(z) \over D(0)} + \overline{b} - 1
\end{equation}
with
\begin{equation}
\label{appBh}
\overline{b} = \left( \xi_{\rm gg}^{\rm 2h}(r,0) 
\over \xi_{\rm dm}^{\rm 2h}(r,0) \right)^{1/2}
\end{equation}
the average bias of galaxies with $L_1  < L < L_2$ at $z=0$, which can
be   obtained   from    the   CLF   using   equations~(\ref{averbias})
and~(\ref{nlm}).

\end{document}